\documentclass[aip,jmp,reprint,amsmath,amssymb,nofootinbib]{revtex4-2}

\usepackage[T1]{fontenc}
\usepackage{bm}
\usepackage{graphicx}
\usepackage{booktabs}
\usepackage{array}
\usepackage{tikz}
\usepackage{enumitem}
\usetikzlibrary{positioning,arrows.meta,fit,backgrounds,calc}
\usepackage{hyperref}
\hypersetup{hidelinks,
  pdftitle={Post-Newtonian Cross Terms in Hierarchical Triple Systems:
    Multiple-Scale Averaging and its Canonical Consistency},
  pdfauthor={Hideyoshi Arakida}}
\graphicspath{{./}}
\definecolor{limc}{RGB}{193,66,66}
\definecolor{kunc}{RGB}{45,95,160}
\definecolor{lic}{RGB}{30,120,90}

\newcommand{\grade}[2]{(#1,\,#2)}
\newcommand{\avt}[1]{\big\langle #1\big\rangle_{t}}
\newcommand{\avF}[1]{\big\langle #1\big\rangle_{F}}
\newcommand{\avo}[1]{\big\langle #1\big\rangle_{1}}
\newcommand{\avtw}[1]{\big\langle #1\big\rangle_{2}}
\newcommand{\Kmq}{{\cal K}_{\rm MQ}}
\newcommand{\M}{\mathcal M}
\newcommand{\Rr}{\mathcal R}
\newcommand{\Ss}{\mathcal S}
\newcommand{\Ww}{\mathcal W}

\setlist{itemsep=1pt,parsep=0pt,topsep=2pt}

\begin{document}

\title{First Post-Newtonian Cross Terms in Hierarchical Triple Systems:
  Multiple-Scale Averaging and its Canonical Consistency}

\author{Hideyoshi Arakida}
\email{arakida.hideyoshi@nihon-u.ac.jp}
\affiliation{College of Engineering, Nihon University,
1 Nakagawara, Tokusada, Tamuramachi, Koriyama, Fukushima 963-8642, Japan}

\date{\today}

\begin{abstract}
We develop a two-stage multiple-scale method for hierarchical
triple systems, using osculating orbital elements and time as
the independent variable.
We examine the disputed cross terms at orders
$\delta\varepsilon^{3/2}$ and $\delta\varepsilon^{7/2}$,
for which the multiple-scale results of Lim and Rodriguez
(LR) disagree with the effective field theory results of
Kuntz, Serra, and Trincherini (KST).
Here, $\delta$ is the first post-Newtonian (1PN) parameter,
and $\varepsilon=a_*/a_2$ ($a_*\simeq a_1$) measures
the orbital hierarchy.
These terms arise from couplings between relativistic
interactions and the tidal field of the third body.
Our results agree with KST: there is no secular term at order
$\delta\varepsilon^{3/2}$, while the
$\delta\varepsilon^{7/2}$ contribution consists of
three parts: a direct term, a first-stage coupling between
the inner binary's center-of-mass motion and the tidal field,
and a second-stage coupling between the 1PN dipole and
the quadrupole.
Their sum reproduces the inner-orbit secular dynamics
generated by KST's magnetic 1PN--quadrupole Hamiltonian,
including its exact mass dependence, for nonsingular
eccentricities and inclinations in the nonresonant,
hierarchical weak-field regime.
The analysis includes mean-anomaly perturbations and
resubstitution of short-period corrections in both directions.
We clarify how the choice of averaging variable affects
the normalization of the mean elements and establish
agreement with canonical perturbation theory.
In particular, osculating and canonical mean elements
give the same inner-orbit secular rates at this order.
\end{abstract}

\maketitle
\section{Introduction}
\label{sec:intro}
Hierarchical triple systems consist of an inner binary with masses $m_1$ and $m_2$ and
a distant tertiary of mass $m_3$, with $a_1/a_2\ll1$
[Fig.~\ref{fig:geom}(a)].
They occur across a wide range of astrophysical environments,
from planetary and stellar systems to compact-object binaries and systems containing
massive black holes.
On secular timescales, angular-momentum exchange between the inner and outer orbits
drives changes in the inner binary's eccentricity and the mutual inclination
through the von Zeipel--Lidov--Kozai (ZLK) mechanism
\cite{vonZeipel1910,Lidov1962,Kozai1962,ItoOhtsuka2020,Naoz2016,Shevchenko2017}.
This evolution may produce extreme eccentricities,
orbital flips, tidal interactions, collisions, and compact-object mergers
\cite{NaozEtAl2013,Naoz2016}.

General relativistic effects further modify the secular dynamics.
Relativistic pericenter precession tends to suppress Newtonian eccentricity excitation
\cite{BlaesLeeSocrates2002,FabryckyTremaine2007,LiuMunozLai2015},
whereas relativistic resonances may instead enhance it
\cite{FordKozinskyRasio2000,NaozKocsisLoebYunes2013,Kuntz2022}.
The resulting high eccentricities accelerate
gravitational-wave-driven orbital decay and merger
\cite{AntoniniPerets2012,SilsbeeTremaine2017,RodriguezAntonini2018,LiuLai2018}.

At first post-Newtonian (1PN) order, relativistic effects include not only
two-body pericenter precession but also genuine three-body interactions
described by the Einstein--Infeld--Hoffmann (EIH) equations
\cite{EinsteinInfeldHoffmann1938}.
We characterize the relativistic and hierarchical expansions by
\begin{equation}
  \delta\equiv\frac{Gm}{c^2a_*},
  \qquad
  \varepsilon\equiv\frac{a_*}{a_2},
  \qquad
  m=m_1+m_2,
  \label{eq:params}
\end{equation}
where $a_*$ is a fixed reference scale comparable to the inner semimajor axis
$a_1$.
The combined expansion generates \emph{cross terms} of order
$\delta\varepsilon^{k/2}$ that couple relativistic dynamics to the
perturbation from the tertiary.
These terms may accumulate over relativistic precession timescales and are
required for a consistent description of the long-term dynamics and
conserved quantities
\cite{NaozKocsisLoebYunes2013,Will2014,Will2014b,LimRodriguez2020,KuntzSerraTrincherini2023}.

Several approaches have been used to derive these cross terms,
but their results do not fully agree.
Naoz \emph{et al.}~\cite{NaozKocsisLoebYunes2013}
doubly averaged the 1PN three-body Hamiltonian at leading
order in $\varepsilon$, while Will \cite{Will2014,Will2014b}
used osculating elements for a circular outer orbit.
Lim and Rodriguez \cite{LimRodriguez2020} (LR) applied
a two-parameter multiple-scale expansion to the EIH equations
using osculating elements and allowing arbitrary eccentricities.
Kuntz, Serra, and Trincherini
\cite{KuntzSerraTrincherini2021,KuntzSerraTrincherini2023}
(KST) derived the complete 1PN--quadrupole cross terms
using nonrelativistic effective field theory and
near-identity transformations of contact elements.
Li, Wu, Younsi, and Li \cite{LiWuYounsiLi2025} (LWYL)
used the ADM three-body Hamiltonian
\cite{OhtaEtAl1974,Schafer1987} together with a von Zeipel transformation.

Two discrepancies arise between LR and KST.
At order $\delta\varepsilon^{3/2}$, LR reported a
``libration'' term [their Eq.~(4.5)] that is absent from the KST and LWYL results.
At order $\delta\varepsilon^{7/2}$, the magnetic-quadrupole
terms of LR [their Eqs.~(A7)--(A11)] differ from those of KST
in both mass dependence and angular structure.
In the examples studied by LR, the libration term also substantially
accelerates octupole-assisted mergers
\cite{Arakida2026PaperII}.

These discrepancies require careful comparison because
the secular equations depend on the definition of the mean orbital elements.
Near-identity transformations between different definitions
may contribute at the same perturbative order as the cross terms.
Meaningful comparisons therefore require a common set of variables.
A consistent derivation must
(A) specify the averaging variables,
(B) include all contributions generated by short-period resubstitution,
and
(C) fix the normalization of the short-period solutions that defines
the mean elements.

In this paper, we develop a two-stage multiple-scale analysis of the
EIH equations that consistently incorporates these three requirements.
We resolve the discrepancies at
$\delta\varepsilon^{3/2}$ and $\delta\varepsilon^{7/2}$
and establish agreement with canonical perturbation theory.
\footnote{A companion paper \cite{Arakida2026PaperII} independently
derives the secular Hamiltonian using Lie--Hori--Deprit
transformations of the ADM Hamiltonian and examines
its implications for orbital evolution.}

Our main findings are as follows.
Time averaging, equivalently mean-anomaly averaging, yields no secular
term at order $\delta\varepsilon^{3/2}$ when the short-period corrections
are normalized to have zero time average.
At order $\delta\varepsilon^{7/2}$, we identify three contributions
(Fig.~\ref{fig:grade}): the direct average of the three-body
EIH acceleration, a first-stage coupling between the
center-of-mass motion of the inner binary and the tidal field, and
a second-stage coupling between the 1PN dipole and the quadrupole.
The last contribution requires resubstitution of the short-period
solutions in both directions.
Their sum reproduces the inner-orbit secular dynamics generated
by the magnetic 1PN--quadrupole Hamiltonian of KST,
including the exact mass dependence of its coefficient
(Sec.~\ref{sec:res}), and agrees with canonical perturbation theory.
The osculating and canonical mean elements yield equivalent
inner-orbit secular equations at this order
(Sec.~\ref{sec:canon}).


The remainder of this paper is organized as follows.
Section~\ref{sec:eq} introduces the EIH equations,
Jacobi coordinates, osculating elements, and the perturbative
order classification.
Section~\ref{sec:ms} develops the two-stage multiple-scale
method, including successive inner- and outer-orbit averaging,
perturbative order counting, and averaging-gauge freedom.
Section~\ref{sec:res} evaluates the cross terms at orders
$\grade{1}{3/2}$, $\grade{1}{5/2}$, and $\grade{1}{7/2}$
and discusses their range of validity.
Section~\ref{sec:canon} compares the results with
canonical perturbation theory and examines the relation
between osculating and canonical mean elements,
the correspondence between resubstitution and Poisson brackets,
and the integrability conditions.
Section~\ref{sec:sum} summarizes the main findings.

\begin{figure*}[tb]
\centering
\resizebox{0.92\linewidth}{!}{\input{tikzA_geom}}
\caption{
  (a) Geometry of a hierarchical triple system.
  The inner binary consists of masses $m_1$ and $m_2$
  with semimajor axis $a_1$, and the tertiary $m_3$
  orbits with semimajor axis $a_2$.
  The vector $\bm R$ points from the Newtonian center of mass
  of the inner binary to $m_3$.
  (b) Orbital angles.
  The reference plane is the invariable plane,
  which lies close to the outer orbital plane (gray).
  The mutual inclination is $\iota$, and the ascending nodes
  differ by $\pi$, giving $\Omega_2=\Omega_1+\pi$.
  The arguments of pericenter $\omega_1$ and $\omega_2$
  are measured from their respective ascending nodes.
  In Secs.~\ref{sec:res} and \ref{sec:canon},
  the outer orbit is held fixed and its plane
  serves as the reference plane (Sec.~\ref{sec:frame}).
}
\label{fig:geom}
\end{figure*}
\section{Equations of Motion and Orbital Elements}
\label{sec:eq}

\subsection{Coordinates and the EIH Equations}

We define the total mass, reduced masses, and symmetric mass
ratio as
$M\equiv m+m_3$, $\mu_1\equiv m_1m_2/m$,
$\mu_2\equiv mm_3/M$, and $\nu\equiv m_1m_2/m^2$.
Following LR, we use the Jacobi coordinates $\bm r$ and $\bm R$,
\begin{equation}
  \bm r\equiv\bm x_1-\bm x_2,\qquad
  \bm R\equiv\bm x_3-\frac{m_1\bm x_1+m_2\bm x_2}{m},
  \label{eq:coords}
\end{equation}
which, in the Newtonian barycentric frame, imply
$\bm x_1=(m_2/m)\bm r-(m_3/M)\bm R$,
$\bm x_2=-(m_1/m)\bm r-(m_3/M)\bm R$, and
$\bm x_3=(m/M)\bm R$.
The Newtonian center of mass of the inner binary therefore moves with
velocity
$\bm V_{\rm cm}=-(m_3/M)\dot{\bm R}$.
This motion contributes to the 1PN three-body interactions and must be retained.

The EIH accelerations are
\cite{EinsteinInfeldHoffmann1938,PoissonWill2014}
\begin{widetext}
  \begin{equation}
    \begin{split}
      \bm a_A={}&-\sum_{B\ne A}\frac{Gm_B}{r_{AB}^2}\bm n_{AB}
      +\frac{1}{c^2}\sum_{B\ne A}\frac{Gm_B}{r_{AB}^2}
      \bigg\{
      \bm n_{AB}\Big[
        4\frac{Gm_B}{r_{AB}}+5\frac{Gm_A}{r_{AB}}
        -v_A^2+4\bm v_A\!\cdot\!\bm v_B\\
        &-2v_B^2+\frac32(\bm n_{AB}\!\cdot\!\bm v_B)^2
        +\sum_{C\ne A,B}\Big(
        4\frac{Gm_C}{r_{AC}}+\frac{Gm_C}{r_{BC}}
        -\frac{Gm_Cr_{AB}}{2r_{BC}^2}
        \bm n_{AB}\!\cdot\!\bm n_{BC}
        \Big)\Big]\\
      &+\big[\bm n_{AB}\!\cdot\!(4\bm v_A-3\bm v_B)\big]
      (\bm v_A-\bm v_B)
      \bigg\}
      -\frac{7}{2c^2}\sum_{B\ne A}\frac{Gm_B}{r_{AB}}
      \sum_{C\ne A,B}\frac{Gm_C}{r_{BC}^2}\bm n_{BC},
    \end{split}
    \label{eq:EIH}
  \end{equation}
\end{widetext}
where $\bm n_{AB}=(\bm x_A-\bm x_B)/r_{AB}$.
These equations follow from the EIH Lagrangian through the Euler--Lagrange
equations \cite{LandauLifshitz1975}.
Writing the Lagrangian as
\begin{equation}
  {\cal L}
  =\sum_A\frac12m_Av_A^2
  +\sum_{A<B}\frac{Gm_Am_B}{r_{AB}}
  +\frac{1}{c^2}{\cal L}_{\rm 1PN}(\bm x,\bm v),
\end{equation}
the Legendre transformation gives
\begin{equation}
  \begin{split}
    \bm p_A
    &=m_A\bm v_A
    +\frac{1}{c^2}
    \frac{\partial{\cal L}_{\rm 1PN}}{\partial\bm v_A},\\
    {\cal H}_{\rm 1PN}(\bm x,\bm p)
    &=-{\cal L}_{\rm 1PN}(\bm x,\bm v)
    \big|_{\bm v_A=\bm p_A/m_A},
  \end{split}
  \label{eq:legendre}
\end{equation}
where ${\cal H}_{\rm 1PN}$ denotes the coefficient of $c^{-2}$
in the Hamiltonian.
At 1PN order, it agrees with the ADM Hamiltonian when expressed
in the corresponding canonical variables
\cite{OhtaEtAl1974,Schafer1987}.

In the zero-total-momentum frame, the momenta
$\bm p$ and $\bm P$, conjugate to $\bm r$ and $\bm R$,
satisfy
\begin{equation}
  \bm p_1=\bm p-\frac{m_1}{m}\bm P,\qquad
  \bm p_2=-\bm p-\frac{m_2}{m}\bm P,\qquad
  \bm p_3=\bm P.
\end{equation}
At Newtonian order,
$\bm p=\mu_1\dot{\bm r}$ and
$\bm P=\mu_2\dot{\bm R}$.
Their 1PN corrections distinguish osculating from contact elements
(Sec.~\ref{sec:canon-mean}).

The relative motion of the inner binary obeys
\begin{equation}
  \ddot{\bm r}
  =-\frac{Gm}{r^3}\bm r+\bm f,\qquad
  \bm f=\bm f_{\rm N}+\frac{1}{c^2}\bm f_{\rm PN},
  \label{eq:relEOM}
\end{equation}
where $\bm f_{\rm N}$ is the Newtonian tidal acceleration and
$\bm f_{\rm PN}$ contains the two-body and three-body 1PN contributions
to $\bm a_1-\bm a_2$.

\subsection{Osculating Elements and Perturbation Equations}

The osculating elements of the inner binary,
$X=(p_1,e_1,\iota_1,\Omega_1,\omega_1)$,
together with the mean anomaly $\M_1$, are the Keplerian elements
associated with the instantaneous position and velocity
$(\bm r,\dot{\bm r})$.
Here,
$p_1=a_1\ell_1^2$, with
$\ell_i\equiv(1-e_i^2)^{1/2}$.

Defining the specific angular momentum
$\bm h=\bm r\times\dot{\bm r}$ and the eccentricity vector
$\bm e_1=\dot{\bm r}\times\bm h/(Gm)-\bm r/r$,
we obtain
\begin{equation}
  \dot{\bm h}=\bm r\times\bm f,\qquad
  \dot{\bm e}_1=
  \frac{\bm f\times\bm h+
    \dot{\bm r}\times(\bm r\times\bm f)}{Gm},
  \label{eq:gaussvec}
\end{equation}
which determine the evolution of the five orbital elements.

Decomposing $\bm f$ into radial, transverse, and normal
components $(\Rr,\Ss,\Ww)$, Gauss's planetary equations
\cite{MurrayDermott1999} are
\begin{equation}
  \begin{split}
    \dot p_1&=\frac{2hr}{Gm}\,\Ss,\\
    \dot e_1&=\frac{1}{h}\Big\{p_1\sin f\,\Rr
      +\big[(p_1+r)\cos f+re_1\big]\Ss\Big\},\\
    \dot\iota_1&=\frac{r\cos\theta_1}{h}\,\Ww,\\
    \dot\Omega_1&=\frac{r\sin\theta_1}{h\sin\iota_1}\,\Ww,\\
    \dot\omega_1&=\frac{1}{he_1}\Big[-p_1\cos f\,\Rr
      +(p_1+r)\sin f\,\Ss\Big]
      -\frac{r\sin\theta_1\cos\iota_1}{h\sin\iota_1}\,\Ww,
  \end{split}
  \label{eq:gauss}
\end{equation}
where $\theta_1=\omega_1+f$ is the argument of latitude,
$f$ is the true anomaly,
$h=|\bm h|=(Gmp_1)^{1/2}$, and $r=|\bm r|$.

The mean anomaly evolves according to
\begin{equation}
  \dot\M_1=n_1(X)+g,\qquad
  g=\frac{b_1}{a_1he_1}
  \Big[(p_1\cos f-2re_1)\Rr
    -(p_1+r)\sin f\,\Ss\Big],
  \label{eq:Mdot}
\end{equation}
where $b_1=a_1\ell_1$ and
\[
  n_1(X)=\left(\frac{Gm}{a_1^3}\right)^{1/2},
  \qquad
  a_1=\frac{p_1}{1-e_1^2}.
\]
We display the argument $X$ explicitly because $n_1$ depends on
the osculating elements through $p_1$ and $e_1$.
In the multiple-scale analysis
(Sec.~\ref{sec:ms-two}),
$X$ is decomposed into slowly varying mean elements
$\tilde X$ and short-period corrections $u$.
The mean motion is then evaluated as $n_1(\tilde X)$,
while its first-order variation
$D_Xn_1\cdot u$ enters the mean-anomaly equation
[Eq.~\eqref{eq:st1per}].

Because $\bm f$ is velocity dependent, we use Gauss's equations
rather than the standard Lagrange planetary equations.
For a velocity-dependent disturbing function, the latter generally
describe contact rather than osculating elements
\cite{EfroimskyGoldreich2003,Efroimsky2005}.

We write the osculating-element equations as
\[
  \dot X=\bm F(X,\M_1;\M_2),
\]
where the outer mean anomaly $\M_2$ enters through
$\bm R$ and $\dot{\bm R}$.
The outer orbit is specified by
$(p_2,e_2,\iota_2,\Omega_2,\omega_2)$,
with mean motion
$n_2=(GM/a_2^3)^{1/2}$.
At the perturbative orders considered here,
the outer orbit may be treated as Keplerian when evaluating
the inner secular rates
(Sec.~\ref{sec:ms-order}).

\paragraph*{Reference frame.}
\label{sec:frame}
We hold the outer orbit fixed and choose its orbital plane
as the reference plane, so that
$\iota_1=\iota$ and $\iota_2=0$.
We retain the convention
$\Omega_2=\Omega_1+\pi$
from the invariable-plane geometry
[Fig.~\ref{fig:geom}(b)],
although the outer ascending node is undefined at $\iota_2=0$.
With this convention,
$\omega_2=\varpi_2-\Omega_2$, where $\varpi_2$
is the longitude of the outer pericenter.

For $G_2\gg G_1$, where $G_i$ denotes the magnitude
of the orbital angular momentum,
the outer orbital plane approaches the invariable plane used by LR.
The two planes differ by an angle of order
$(G_1/G_2)\sin\iota$.
This small rotation does not affect the vector form
of the secular equations.

The derivatives $D_X$ are evaluated with the outer orbit,
including $\varpi_2$, held fixed.
Since
$\omega_2=\varpi_2-\Omega_1-\pi$,
differentiation with respect to $\Omega_1$
at fixed $\varpi_2$ is represented by
\[
  \left.\frac{\partial}{\partial\Omega_1}\right|_{\varpi_2}
  =
  \left.\frac{\partial}{\partial\Omega_1}\right|_{\omega_2}
  -
  \frac{\partial}{\partial\omega_2}.
\]
We evaluate only the five inner-orbit rates,
which we refer to as the \emph{inner sector}.
Outer-orbit back-reaction enters at higher order and is not
included here
(Sec.~\ref{sec:ms-order}).
Vector expressions such as Eqs.~\eqref{eq:dS}
and \eqref{eq:Kmq} are independent of this
reference-frame choice.

\subsection{Perturbative Orders}

We denote a term in the secular equations by
$\grade{n}{k/2}$ if it scales as
$n_*\delta^n\varepsilon^{k/2}$, where
$n_*=(Gm/a_*^3)^{1/2}$,
with mass-ratio factors treated separately.
The characteristic velocity ratio scales as
$\dot R/\dot r\sim\varepsilon^{1/2}(M/m)^{1/2}$,
the tidal perturbation as $\varepsilon^3$,
and the mean-motion ratio as
$n_2/n_1=\varepsilon^{3/2}(M/m)^{1/2}$.
These scalings naturally generate half-integer orders.

Table~\ref{tab:orders} summarizes the perturbations relevant here.
For numerical order separation, we fix $a_1$ and set
$a_2=a_1\lambda^{-2}$.
A term of order $\grade{1}{k/2}$ then scales as
$\lambda^k c^{-2}$, allowing the coefficients
at different orders to be extracted separately
(Appendix~\ref{app:num}).

\begin{table}[tb]
  \caption{
    Perturbative orders and contributions after
    inner-orbit averaging.
    ``Odd'' denotes terms proportional to $m_1-m_2$
    that are odd in the outer position or velocity.
    The zeros at $\grade{1}{\frac12}$ and
    $\grade{1}{1}$, and the absence of a
    $\grade{1}{\frac32}$ contribution, are exact
    (Sec.~\ref{sec:ms-order}).
  }
  \label{tab:orders}
  \centering\small
  \begin{tabular}{@{}lll@{}}
    \toprule
    Order & Origin & After inner average\\
    \midrule
    $\grade{0}{3}$, $\grade{0}{4}$
    & Newtonian quadrupole, octupole & ZLK, octupole\\
    $\grade{1}{0}$
    & Two-body 1PN & Apsidal precession\\
    $\grade{1}{\frac12}$
    & Center-of-mass motion, odd & Zero\\
    $\grade{1}{1}$
    & Outer velocity squared, binding energy & Zero\\
    $\grade{1}{\frac32}$
    & None (vanishes identically) & ---\\
    $\grade{1}{2}$
    & 1PN dipole, odd & Periodic in $\M_2$\\
    $\grade{1}{\frac52}$
    & de Sitter & Secular\\
    $\grade{1}{3}$, $\grade{1}{\frac72}$
    & Electric, magnetic 1PN--quadrupole & Secular\\
    \bottomrule
  \end{tabular}
\end{table}

\section{The Multiple-Scale Method}
\label{sec:ms}

\subsection{Two-Parameter Expansion and Timescales}
\label{sec:ms-exp}

The perturbation equations in Sec.~\ref{sec:eq}
involve two small parameters, $\delta$ and $\varepsilon$.
The Newtonian tidal acceleration scales as
$(m_3/m)(r/R)^3\sim\varepsilon^3$ relative to
the inner Keplerian acceleration, with mass-ratio
factors treated separately.
Its multipole expansion contains integer powers
of $\varepsilon$.

The 1PN acceleration scales as $\delta$ relative
to the inner Keplerian acceleration.
Its three-body terms depend on
$r/R\sim\varepsilon$ and on the velocity ratio
$V/v\sim\varepsilon^{1/2}(M/m)^{1/2}$,
which introduces half-integer powers of
$\varepsilon$.

We expand the orbital-element rates and
mean-anomaly perturbation as
\begin{equation}
  \begin{split}
    \bm F&=n_*\sum_{n=0}^{1}\sum_k
    \delta^n\varepsilon^{k/2}
    \bm F^{\grade{n}{k/2}}(X,\M_1;\M_2),\\
    g&=n_*\sum_{n=0}^{1}\sum_k
    \delta^n\varepsilon^{k/2}
    g^{\grade{n}{k/2}},
  \end{split}
  \label{eq:Fexp}
\end{equation}
where $n_*=(Gm/a_*^3)^{1/2}$, and
$\delta$ and $\varepsilon$ are the fixed parameters
defined in Eq.~\eqref{eq:params}.
The expansion coefficients contain the dependence
on the orbital elements, anomalies, and mass ratios,
including factors such as $n_1(X)/n_*$ and
$(a_1/a_*)^j$.
Derivatives $D_X$ act on these coefficients with
the reference parameters and orbital anomalies held fixed.

The Newtonian expansion begins at quadrupole order
($k=6$), followed by octupole order ($k=8$).
The 1PN expansion begins with the two-body contribution
($k=0$), followed by the three-body terms listed in
Table~\ref{tab:orders}.
Terms of order $\delta^2$ and higher are neglected.

The dynamics involves the inner and outer orbital periods
as well as longer secular timescales.
The mean motions satisfy
\begin{equation}
  n_2=\kappa\,\varepsilon^{3/2}n_*,
  \qquad
  \frac{n_2}{n_1}
  =\kappa\,\varepsilon^{3/2}
  \left(\frac{a_1}{a_*}\right)^{3/2},
  \qquad
  \kappa\equiv\left(\frac{M}{m}\right)^{1/2}.
  \label{eq:n2n1}
\end{equation}
The first relation is exact for the fixed Keplerian
outer orbit.
We use $a_1\simeq a_*$ for order estimates.

The characteristic secular timescales are
$(\delta n_1)^{-1}$ for 1PN apsidal precession
and $(\varepsilon^3n_1)^{-1}$ for the ZLK mechanism,
apart from mass-ratio and eccentricity factors.

The method of multiple scales
\cite{Nayfeh1973,KevorkianCole1996,SandersVerhulstMurdock2007}
separates the fast orbital angles from the slowly
evolving mean elements.
We first average over the inner mean anomaly $\M_1$,
holding the outer orbit fixed, and then over the
outer mean anomaly $\M_2$.

We introduce the near-identity transformations
\begin{equation}
  X=\tilde X+u(\tilde X,\phi_1;\M_2),\quad
  \M_1=\phi_1+w(\tilde X,\phi_1;\M_2),\quad
  \tilde X=\bar X+U(\bar X,\M_2),
  \label{eq:ansatz}
\end{equation}
where $u$ and $w$ are $2\pi$-periodic in
$\phi_1$, and $U$ is $2\pi$-periodic in $\M_2$.
Here, $\tilde X$ and $\bar X$ denote the
inner-averaged and doubly averaged elements,
respectively.

The doubly averaged elements obey
\begin{equation}
  \dot{\bar X}
  =n_*\sum_{n=0}^{1}\sum_k
  \delta^n\varepsilon^{k/2}
  \bm V^{\grade{n}{k/2}}(\bar X).
  \label{eq:Vexp}
\end{equation}
Our goal is to determine the 1PN secular coefficients
$\bm V^{\grade{1}{k/2}}$, particularly the cross terms
that couple relativistic and third-body perturbations.

Each averaging stage requires the secular evolution
to be slow compared with the orbital frequency being
averaged.
For the outer averaging, the relevant ratios are
\begin{equation}
  \frac{3\delta n_1}{\ell_1^2n_2},
  \qquad
  \frac{m_3}{m}
  \frac{\varepsilon^3n_1}{\ell_2^3n_2}.
\end{equation}
For $a_1\simeq a_*$, these become
$3\delta/(\kappa\varepsilon^{3/2}\ell_1^2)$ and
$(m_3/m)\varepsilon^{3/2}/(\kappa\ell_2^3)$,
respectively.
Both must be small.
In particular, 1PN precession must remain slow
relative to the outer orbital motion.
For $m_3\gg m$, the second ratio approaches
$\kappa\varepsilon^{3/2}/\ell_2^3$, requiring
a stronger hierarchy (Sec.~\ref{sec:valid}).
Under these conditions, we treat $\delta$ and
$\varepsilon$ as independent formal expansion
parameters and retain terms through first order
in $\delta$.

\subsection{Two-Stage Time Averaging}
\label{sec:ms-two}

\paragraph*{First stage: Inner-orbit averaging.}
Holding the outer orbit fixed, the inner angle evolves as
$\dot\phi_1=n_1(\tilde X)+\avo{g}+\cdots$.
Here, $\avo{\cdot}$ denotes averaging over $\phi_1$,
equivalent to time averaging along the unperturbed
inner Keplerian orbit with $\tilde X$ and $\M_2$ fixed.

We expand the periodic corrections and
inner-averaged rates as
\begin{equation}
  \begin{split}
    u&=\sum_{n,k}\delta^n\varepsilon^{k/2}
    u^{\grade{n}{k/2}}+\cdots,\\
    w&=\sum_{n,k}\delta^n\varepsilon^{k/2}
    w^{\grade{n}{k/2}}+\cdots,\\
    \dot{\tilde X}
    &=n_*\sum_{n,k}\delta^n\varepsilon^{k/2}
    \bm V_1^{\grade{n}{k/2}}(\tilde X;\M_2)
    +\cdots.
  \end{split}
  \label{eq:st1exp}
\end{equation}

Substituting Eqs.~\eqref{eq:ansatz} and
\eqref{eq:st1exp} into the perturbation equations,
we use
\begin{equation*}
  \frac{d}{dt}
  =\dot\phi_1\,\partial_{\phi_1}
  +\dot{\tilde X}\cdot D_{\tilde X}
  +n_2\partial_{\M_2}.
\end{equation*}
Expanding about $(\tilde X,\phi_1)$ and collecting
terms at each perturbative order yields the
periodic corrections.

At first order,
\begin{equation}
  \begin{split}
    n_1\frac{\partial u^{\grade{n}{k/2}}}
    {\partial\phi_1}
    &=n_*\left[
      \bm F^{\grade{n}{k/2}}
      -\avo{\bm F^{\grade{n}{k/2}}}
    \right],\\
    n_1\frac{\partial w^{\grade{n}{k/2}}}
    {\partial\phi_1}
    &=n_*\left[
      g^{\grade{n}{k/2}}
      -\avo{g^{\grade{n}{k/2}}}
    \right]
    +D_Xn_1\cdot u^{\grade{n}{k/2}},
  \end{split}
  \label{eq:st1per}
\end{equation}
where $n_1=n_1(\tilde X)$ and
$\bm V_1^{\grade{n}{k/2}}
=\avo{\bm F^{\grade{n}{k/2}}}$.
The last term in the equation for $w$ accounts for
the change in the inner mean motion caused by
short-period variations in $a_1$.

At second order, resubstitution of the periodic
corrections generates
\begin{equation}
  \avo{
    D_X\bm F^{\grade{n}{k/2}}
    \cdot u^{\grade{n'}{k'/2}}
    +\partial_{\M_1}\bm F^{\grade{n}{k/2}}\,
    w^{\grade{n'}{k'/2}}
  },
  \label{eq:st1sec}
\end{equation}
at order
$\delta^{n+n'}\varepsilon^{(k+k')/2}$,
summed over ordered pairs.
Here, $D_X$ acts on the expansion coefficients
at fixed $\M_1$ and $\M_2$.

The contribution
$\avo{D_{\tilde X}u\cdot\bm V_1}$
vanishes under the normalization $\avo{u}=0$
(Sec.~\ref{sec:ms-gauge}).
The second term in Eq.~\eqref{eq:st1sec}
is the periodic mean-anomaly, or ``clock,''
correction.
LR incorporate this effect through the
true-anomaly evolution equation.

The term $n_2\partial_{\M_2}u$ is smaller than
$n_1\partial_{\phi_1}u$ by a factor of order
$\kappa\varepsilon^{3/2}$ and must therefore
be retained at the corresponding higher orders.

For a 1PN perturbation of order $\grade{1}{a}$
and a Newtonian perturbation of order
$\grade{0}{b}$, the first-stage cross term is
\begin{equation}
  \begin{split}
    \mathsf S_1^{\grade{1}{a}\otimes\grade{0}{b}}
    =\avo{
      &D_X\bm F^{\grade{1}{a}}
      \cdot u^{\grade{0}{b}}
      +D_X\bm F^{\grade{0}{b}}
      \cdot u^{\grade{1}{a}}\\
      &+\partial_{\M_1}\bm F^{\grade{1}{a}}\,
      w^{\grade{0}{b}}
      +\partial_{\M_1}\bm F^{\grade{0}{b}}\,
      w^{\grade{1}{a}}
    }.
  \end{split}
  \label{eq:S1k}
\end{equation}
The superscript
$\grade{1}{a}\otimes\grade{0}{b}$
denotes the second-order coupling between
the two perturbations.
Each term is bilinear, with the rate of one
perturbation multiplied by the short-period
correction generated by the other.
Thus, $\otimes$ denotes the pairing of both
orderings, not a tensor product.
The resulting contribution is of order
$\grade{1}{a+b}$.

Summing over all such pairs gives
\begin{equation}
  \begin{split}
    \mathsf S_1=\avo{
      &D_X\bm F_{\rm N}\cdot u_{\rm PN}
      +D_X\bm F_{\rm PN}\cdot u_{\rm N}
      +\partial_{\M_1}\bm F_{\rm N}\,w_{\rm PN}
      +\partial_{\M_1}\bm F_{\rm PN}\,w_{\rm N}
    },
  \end{split}
  \label{eq:S1}
\end{equation}
where
$u=u_{\rm N}+c^{-2}u_{\rm PN}$
and
$w=w_{\rm N}+c^{-2}w_{\rm PN}$.

The resulting inner-averaged field is
\begin{equation*}
  \dot{\tilde X}
  =\bar{\bm F}(\tilde X;\M_2)
  =n_*\sum_{n,k}\delta^n\varepsilon^{k/2}
  \bar{\bm F}^{\grade{n}{k/2}},
\end{equation*}
with
\begin{equation}
  \begin{split}
    \bar{\bm F}^{\grade{0}{k/2}}
    &=\avo{\bm F^{\grade{0}{k/2}}}+\cdots,\\
    \bar{\bm F}^{\grade{1}{k/2}}
    &=\avo{\bm F^{\grade{1}{k/2}}}
    +\sum_{a+b=k/2}
    \mathsf S_1^{\grade{1}{a}\otimes\grade{0}{b}}.
  \end{split}
  \label{eq:Fbar}
\end{equation}
The ellipsis denotes higher-order Newtonian
products, beginning at $\grade{0}{6}$.
Because Newtonian perturbations begin at
$\grade{0}{3}$, first-stage
1PN--Newtonian products first appear at
$\grade{1}{3}$.

Through order $\grade{1}{2}$, the nonzero
inner-averaged 1PN terms are
$\bar{\bm F}^{\grade{1}{0}}$, describing
apsidal precession, and the 1PN dipole
$\bar{\bm F}^{\grade{1}{2}}$.
The former is independent of $\M_2$,
whereas the outer-orbit average of the latter
vanishes.
The inner-averaged coefficients at
$\grade{1}{\frac12}$, $\grade{1}{1}$,
and $\grade{1}{\frac32}$ vanish
(Table~\ref{tab:orders};
Sec.~\ref{sec:ms-order}).

\paragraph*{Second stage: Outer-orbit averaging.}
We next average
$\dot{\tilde X}=\bar{\bm F}(\tilde X;\M_2)$
using
$\tilde X=\bar X+U(\bar X,\M_2)$
and $\dot\M_2=n_2$.
Since $n_2=n_*\kappa\varepsilon^{3/2}$,
the periodic corrections acquire the factor
$n_*/n_2=\kappa^{-1}\varepsilon^{-3/2}$.
Thus,
\begin{equation}
  \begin{split}
    U&=\sum_{n,k}
    \delta^n\varepsilon^{(k-3)/2}
    \kappa^{-1}U^{\grade{n}{k/2}},\\
    \frac{\partial U^{\grade{n}{k/2}}}
    {\partial\M_2}
    &=\bar{\bm F}^{\grade{n}{k/2}}
    -\avtw{\bar{\bm F}^{\grade{n}{k/2}}},
  \end{split}
  \label{eq:st2}
\end{equation}
where $\avtw{\cdot}$ denotes time averaging
over the outer orbit at fixed $\bar X$.
Because $n_2$ is independent of $\bar X$,
no mean-motion correction analogous to that
in Eq.~\eqref{eq:st1per} appears.

The Newtonian quadrupole and 1PN dipole
periodic corrections scale as
$\varepsilon^{3/2}$ and
$\delta\varepsilon^{1/2}$, respectively,
apart from mass-ratio factors.
At first order, the secular coefficients are
$\avtw{\bar{\bm F}^{\grade{n}{k/2}}}$.
At second order, a 1PN term of order
$\grade{1}{a}$ and a Newtonian term of order
$\grade{0}{b}$ generate
\begin{equation}
  \begin{split}
    \mathsf S_2^{\grade{1}{a}\otimes\grade{0}{b}}
    =\kappa^{-1}\avtw{
      &D_X\bar{\bm F}^{\grade{0}{b}}
      \cdot U^{\grade{1}{a}}
      +D_X\bar{\bm F}^{\grade{1}{a}}
      \cdot U^{\grade{0}{b}}
    },
  \end{split}
  \label{eq:S2k}
\end{equation}
at order $\grade{1}{a+b-\frac32}$,
where $D_X$ is evaluated at fixed $\M_2$.

Summing over all pairs gives
\begin{equation}
  \mathsf S_2=\avtw{
    D_X\bar{\bm F}_{\rm N}\cdot U_{\rm PN}
    +D_X\bar{\bm F}_{\rm PN}\cdot U_{\rm N}
  },
  \label{eq:S2}
\end{equation}
where the subscripts ${\rm N}$ and ${\rm PN}$
denote Newtonian and 1PN contributions,
respectively.

Corrections involving the outer orbital
variables enter the inner secular equations
only at the orders identified below.

\paragraph*{Assembly.}
Collecting terms of order
$\delta\varepsilon^{k/2}$ gives
\begin{equation}
  \begin{split}
    \bm V^{\grade{1}{k/2}}
    ={}&\mathsf D^{\grade{1}{k/2}}
    +\sum_{a+b=k/2}
    \avtw{\mathsf S_1^{\grade{1}{a}\otimes\grade{0}{b}}}
    +\sum_{a+b=(k+3)/2}
    \mathsf S_2^{\grade{1}{a}\otimes\grade{0}{b}}
    +\cdots,
  \end{split}
  \label{eq:assembly}
\end{equation}
where
$\mathsf D^{\grade{1}{k/2}}
=\avtw{\avo{\bm F^{\grade{1}{k/2}}}}$
is the direct contribution.
The ellipsis denotes higher-order products,
whose relevance is assessed in
Sec.~\ref{sec:ms-order}.

Using Eqs.~\eqref{eq:S1} and \eqref{eq:S2},
the 1PN secular rates have the schematic form
\[
  \dot{\bar X}_{\rm PN}
  =\mathsf D+\avtw{\mathsf S_1}+\mathsf S_2,
\]
with
$\mathsf D=\avtw{\avo{\bm F_{\rm PN}}}$.

Both stages require resubstitution in both directions:
1PN-generated periodic corrections enter Newtonian
rates, while Newtonian-generated corrections enter
1PN rates.
Together with the direct contribution, these terms
are required to recover the Hamiltonian secular
dynamics
(Sec.~\ref{sec:canon-lie}).

\subsection{Order Counting}
\label{sec:ms-order}

A product of $j$ perturbations of orders
$\grade{n_i}{k_i/2}$ contributes at
\begin{equation*}
  \grade{\sum n_i}{\sum k_i/2+s},
\end{equation*}
where $s$ is the sum of the order shifts
associated with the $j-1$ periodic corrections.
Corrections to the inner variables
$(X,\M_1)$ introduce no shift in the first stage
and a shift of $-\frac32$ in the second stage,
because
$n_*/n_2\propto\varepsilon^{-3/2}$.
Corrections to the outer orbital variables are
suppressed by
$G_1/G_2\sim\varepsilon^{1/2}$
relative to inner-variable corrections.
Their shifts are therefore $+\frac12$ and $-1$
in the first and second stages, respectively,
with mass-ratio and eccentricity factors treated
separately.

For 1PN cross terms, exactly one factor has order
$\grade{1}{a}$, while the remaining Newtonian
factors have integer orders $b\ge3$.

\paragraph*{Lower orders.}
The 1PN relative acceleration contains no term at
$\grade{1}{\frac32}$:
the coefficient of $\lambda^3$ vanishes identically,
as verified by an exact SymPy calculation with
symbolic masses
\cite{ArakidaCodePaperI}.
Odd powers of $\lambda$ require odd powers of the
outer velocity
$\dot{\bm R}\propto\lambda$.
Through order $\lambda^3$, the outer orbit enters
$\bm f_{\rm PN}$ through the center-of-mass velocity,
at most quadratically, and through the
velocity-independent potential
$Gm_3/R\propto\lambda^2$.
The remaining $m_3$-dependent terms begin at
${\cal O}(\lambda^4)$.

At $\grade{1}{\frac12}$ and $\grade{1}{1}$,
the inner-orbit averages of the
angular-momentum, eccentricity-vector, and
Keplerian-energy rates vanish for arbitrary
masses, outer velocities, and inner eccentricities.
These cancellations follow from exact contour
integration using
$e_1=2x/(1+x^2)$ and $z=e^{iE_1}$,
with residues at $z=0$ and $z=x$.

Consequently,
\begin{equation*}
  \bar{\bm F}^{\grade{1}{1/2}}
  =\bar{\bm F}^{\grade{1}{1}}
  =\bar{\bm F}^{\grade{1}{3/2}}=0
\end{equation*}
identically in $(\tilde X,\M_2)$.
Their second-stage periodic corrections and
derivatives therefore vanish.

The corresponding Hamiltonian result,
$\avo{{\cal H}^{\grade{1}{1/2}}}=0$, follows from
\begin{equation*}
  \begin{split}
    \avo{\frac{\dot{\bm r}}{r}}
    &=\avo{
      \frac{(\bm r\cdot\dot{\bm r})\bm r}{r^3}
    },\\
    \avo{|\dot{\bm r}|^2\dot{\bm r}}
    &=2Gm\avo{\frac{\dot{\bm r}}{r}},
  \end{split}
\end{equation*}
whose contributions cancel with the exact
mass coefficients.

\paragraph*{Census of second-order products.}
Table~\ref{tab:census} lists the
1PN--Newtonian products that may contribute at
$\grade{1}{\frac32}$,
$\grade{1}{\frac52}$, or
$\grade{1}{\frac72}$,
including corrections to the outer variables.

At $\grade{1}{\frac72}$, only two products contribute in addition
to the direct term: the first-stage coupling
$\grade{1}{\frac12}\otimes\grade{0}{3}$
and the second-stage coupling
$\grade{1}{2}\otimes\grade{0}{3}$.

The remaining candidates vanish because
(i) the inner averages at
$\grade{1}{\frac12}$ and $\grade{1}{1}$ vanish;
(ii) the two-body term
$\bar{\bm F}^{\grade{1}{0}}$
is independent of $\M_2$ and of the outer variables,
so that $U^{\grade{1}{0}}=0$, while its products
with Newtonian corrections reduce to
$D_X\bar{\bm F}^{\grade{1}{0}}\cdot\avtw{U}=0$
in the time gauge
[Eqs.~\eqref{eq:timegauge} and \eqref{eq:32}];
or (iii) the required perturbation is absent at
$\grade{0}{\frac72}$ or
$\grade{1}{\frac32}$.

\begin{table}[tb]
  \caption{
    Second-order 1PN--Newtonian products
    at the indicated perturbative orders.
    ``Inner'' and ``outer'' identify the
    corrected variables, and $s$ is the order shift.
    Reasons (i)--(iii) are given in the text.
  }
  \label{tab:census}
  \centering\small
  \begin{tabular}{@{}llll@{}}
    \toprule
    Order & Stage, correction ($s$) & Product & Result\\
    \midrule
    $\grade{1}{\frac32}$
      & Direct
      & $\grade{1}{\frac32}$ & Absent (iii)\\
      & 2, inner ($-\frac32$)
      & $\grade{1}{0}\otimes\grade{0}{3}$ & 0 (ii)\\
    \midrule
    $\grade{1}{\frac52}$
      & Direct
      & $\grade{1}{\frac52}$ & de Sitter\\
      & 2, inner ($-\frac32$)
      & $\grade{1}{1}\otimes\grade{0}{3}$ & 0 (i)\\
      &
      & $\grade{1}{0}\otimes\grade{0}{4}$ & 0 (ii)\\
      & 2, outer ($-1$)
      & $\grade{1}{\frac12}\otimes\grade{0}{3}$ & 0 (i)\\
    \midrule
    $\grade{1}{\frac72}$
      & Direct
      & $\grade{1}{\frac72}$ & $\mathsf D$\\
      & 1, inner ($0$)
      & $\grade{1}{\frac12}\otimes\grade{0}{3}$
      & $\avtw{\mathsf S_1}$\\
      & 1, outer ($+\frac12$)
      & $\grade{1}{0}\otimes\grade{0}{3}$ & 0 (ii)\\
      & 2, inner ($-\frac32$)
      & $\grade{1}{2}\otimes\grade{0}{3}$ & $\mathsf S_2$\\
      &
      & $\grade{1}{1}\otimes\grade{0}{4}$ & 0 (i)\\
      &
      & $\grade{1}{0}\otimes\grade{0}{5}$ & 0 (ii)\\
      & 2, outer ($-1$)
      & $\grade{1}{\frac12}\otimes\grade{0}{4}$ & 0 (i)\\
      &
      & $\grade{1}{\frac32}\otimes\grade{0}{3}$
      & Absent (iii)\\
    \bottomrule
  \end{tabular}
\end{table}

\paragraph*{Higher-order products.}
Products of three perturbations do not contribute
through the orders considered here.
In the first stage, their lowest possible order is
$\grade{1}{a+b+c}$, with $b,c\ge3$.
In the second stage, two periodic corrections produce
shifts $s=-3$, $-\frac52$, or $-2$, so the lowest
possible order is
$\grade{1}{a+b+c-3}$.

A third-order product can reach
$\grade{1}{\frac72}$ only in two cases.
The first has $s=-3$, $a=\frac12$, and $b=c=3$,
but vanishes by (i).
The second has $s=-\frac52$, $a=0$, and $b=c=3$,
with one inner and one outer correction.

In the second case, the 1PN factor is
$\bar{\bm F}^{\grade{1}{0}}$.
Its first derivative multiplies the time average
of a second-order correction, which vanishes in
the time gauge.
Its second derivative involves only corrections
to $(p_1,e_1)$, whose orders cannot combine to
produce $\grade{1}{\frac72}$.

Thus, third-order products do not contribute at
the target orders.
First-stage cross terms, beginning at
$\grade{1}{3}$, enter subsequent second-stage
products only at $\grade{1}{\frac92}$ or higher.
The secular evolution of the outer orbit itself
is not considered here.

\begin{figure}[tb]
  \centering
  \resizebox{\columnwidth}{!}{\begin{tikzpicture}[x=8.6mm,y=9mm,font=\scriptsize,>=Stealth]
\def\R{1.55}
\def\hw{0.44}
\def\hh{0.30}
\foreach \k in {0,...,9}{
 \foreach \n in {0,\R}{
  \draw[black!15] (\k-\hw,\n-\hh) rectangle (\k+\hw,\n+\hh);}}
\foreach \k/\lab in {0/{$0$},1/{$\frac12$},2/{$1$},3/{$\frac32$},4/{$2$},
           5/{$\frac52$},6/{$3$},7/{$\frac72$},8/{$4$},9/{$\frac92$}}
 \node[black!70] at (\k,-0.62) {\lab};
\node[black!70] at (4.5,-1.10) {$k/2$ (power of $\varepsilon$)};
\node[black!70,anchor=east] at (-0.55,0) {$n=0$};
\node[black!70,anchor=east] at (-0.55,\R) {$n=1$};
\newcommand{\cell}[4]{%
 \draw[#4] (#1-\hw,#2-\hh) rectangle (#1+\hw,#2+\hh);
 \node at (#1,#2) {#3};}
\cell{6}{0}{Q}{draw=black,fill=black!8,line width=0.6pt}
\cell{8}{0}{O}{draw=black,fill=black!8,line width=0.6pt}
\cell{0}{\R}{1PN}{draw=black,fill=black!8,line width=0.6pt}
\cell{1}{\R}{CM}{draw=black,dashed,line width=0.6pt}
\cell{4}{\R}{dip.}{draw=black,dashed,line width=0.6pt}
\cell{3}{\R}{$0$}{draw=black,line width=1.1pt}
\cell{5}{\R}{dS}{draw=black,fill=black!8,line width=0.6pt}
\cell{7}{\R}{MQ}{draw=black,line width=1.1pt}
\draw[->,black!60] (0,\R+\hh) to[bend left=28] node[above,black]{$\otimes$Q, stage 2} (3,\R+\hh);
\draw[->,black!60] (1,\R-\hh) to[bend right=12] node[below=1pt,black,fill=white,inner sep=1pt]{$\otimes$Q, stage 1} (7,\R-\hh);
\draw[->,black!60] (4,\R+\hh) to[bend left=28] node[above,black]{$\otimes$Q, stage 2} (7,\R+\hh);
\end{tikzpicture}}
  \caption{
    Perturbative orders $\grade{n}{k/2}$
    of individual terms and cross terms (bold).
    Gray denotes terms that remain after the inner average;
    dashed lines indicate terms with zero inner average.
    Arrows show candidate couplings at
    $\grade{1}{3/2}$ and $\grade{1}{7/2}$.
    ``$\otimes$Q, stage~$s$'' denotes coupling
    with the Newtonian quadrupole
    Q $=\grade{0}{3}$ at averaging stage $s$.
    The inner-variable order shifts are $0$
    and $-\frac32$ at stages 1 and 2, respectively.
    CM: center-of-mass motion;
    dip.: 1PN dipole;
    dS: de Sitter;
    MQ: magnetic quadrupole.
  }
  \label{fig:grade}
\end{figure}

\subsection{Angle Variables and Gauge Freedom}
\label{sec:ms-gauge}

The periodic corrections in
Eqs.~\eqref{eq:st1per} and \eqref{eq:st2}
are determined only up to functions of the
slow variables.
For example, the transformation
$U\to U+\gamma(\bar X)$ corresponds, to first
order, to $\bar X\to\bar X-\gamma$.
The secular vector field then changes by
\begin{equation}
  \Delta\dot{\bar X}
  =D_X\avtw{\bar{\bm F}}\cdot\gamma
  -D_X\gamma\cdot\avtw{\bar{\bm F}},
  \label{eq:gauge}
\end{equation}
which is the commutator of $\gamma$ and
$\avtw{\bar{\bm F}}$.
The first term arises from resubstitution,
whereas the second accounts for the evolution
of $\gamma$ along the secular flow.

Thus, the secular equations depend on the
definition of the mean elements.
Because $\avtw{\bar{\bm F}}$ contains both
the Newtonian quadrupole and 1PN apsidal-precession
terms, shifts of order $\varepsilon^{3/2}$ or
$\delta$ may modify the cross terms.

We fix this freedom by imposing zero
\emph{time} averages:
\begin{equation}
  \avo{u}=\avo{w}=0,\qquad
  \avtw{U}=0.
  \label{eq:timegauge}
\end{equation}
This convention is consistent with the
zero-average normalization of generating
functions in Lie--Hori--Deprit canonical
perturbation theory
\cite{Hori1966,Deprit1969}
and facilitates comparison with canonical
mean elements
(Sec.~\ref{sec:canon}).

Uniform averaging over an orbital angle
$\varphi$, such as the eccentric anomaly $E_2$
or true anomaly $F$, generally differs from
time averaging:
\begin{equation}
  \langle Y\rangle_\varphi
  =\frac{1}{2\pi}\int_0^{2\pi}Y\,d\varphi
  =\Big\langle
    Y\frac{d\varphi}{d\M_2}
  \Big\rangle_t.
  \label{eq:measure}
\end{equation}
Table~\ref{tab:measure} illustrates these
differences.

\begin{table}[tb]
  \caption{
    Orbital averages of $\cos(kF)$ over three
    orbital angles.
    Here $k$ is a positive integer,
    $\ell_2=\sqrt{1-e_2^2}$,
    $x=(1-\ell_2)/e_2$, and
    $C_k=(-1)^k(1+k\ell_2)x^k$.
    All three averages of $\sin(kF)$ vanish.
  }
  \label{tab:measure}
  \centering\small
  \begin{tabular}{@{}lccc@{}}
    \toprule
    Angle $\varphi$ &
    $d\varphi/d\M_2$ &
    $\langle\cos(kF)\rangle_\varphi$ &
    $\langle\cos(2F)\rangle_\varphi$\\
    \midrule
    $\M_2$ (time) & $1$ & $C_k$ &
    $\dfrac{(1-\ell_2)(1+2\ell_2)}{1+\ell_2}$\\[6pt]
    $E_2$ &
    $\dfrac{1+e_2\cos F}{\ell_2^2}$ &
    $(-x)^k$ &
    $\dfrac{1-\ell_2}{1+\ell_2}$\\[6pt]
    $F$ &
    $\dfrac{(1+e_2\cos F)^2}{\ell_2^3}$ &
    $0$ & $0$\\
    \bottomrule
  \end{tabular}
\end{table}

Imposing $\langle U\rangle_\varphi=0$
therefore defines a different gauge, with
\[
  \gamma^\varphi
  =\avt{U}-\langle U\rangle_\varphi.
\]

Using $\varphi$ as the independent variable
also introduces a periodic clock correction
$\bar V\,\Delta t_\varphi$, even when the
secular rate $\bar V$ is constant, where
$\Delta t_\varphi=t-\varphi/n_2$.
For the true and eccentric anomalies,
\begin{equation*}
  \Delta t_F=n_2^{-1}(\M_2-F),
  \qquad
  \Delta t_E=-n_2^{-1}e_2\sin E_2,
\end{equation*}
respectively, for a common time origin.

Secular equations obtained in different
angle gauges are related by
Eq.~\eqref{eq:gauge}.
All periodic corrections, including the
clock terms, must therefore be treated
consistently before the results are compared.

At quadrupole order, the gauge shift between
eccentric-anomaly and time averaging vanishes.
Using
$d\M_2=(1-e_2\cos E_2)\,dE_2$,
integration by parts gives
\begin{equation*}
  \avt{Y}-\langle Y\rangle_E
  =e_2\langle Y'\sin E_2\rangle_E,
\end{equation*}
where the prime denotes differentiation
with respect to $E_2$.

For the quadrupole periodic correction $Y=U$,
\begin{equation*}
  Y'=\frac{1}{n_2}
  \left(
    \bar{\bm F}-\avtw{\bar{\bm F}}
  \right)\frac{R}{a_2}.
\end{equation*}
The quadrupole field depends on the outer
separation through $R^{-3}$ and on its direction
through terms containing
$\hat{\bm N}\hat{\bm N}$.

Using
\begin{equation*}
  R^{-2}\sin E_2\,dE_2
  =\frac{\sin F}{a_2^2\ell_2^2}\,dF,
\end{equation*}
the first contribution vanishes because
$\hat{\bm N}\hat{\bm N}$ contains only the
zeroth and second harmonics of $F$:
\begin{equation*}
  \int_0^{2\pi}
  \hat{\bm N}\hat{\bm N}\sin F\,dF=0.
\end{equation*}
The remaining contribution also vanishes because
\[
  \int_0^{2\pi}R\sin E_2\,dE_2=0.
\]

Thus,
$\avt{U}=\langle U\rangle_E$
at quadrupole order.
The two gauges generally differ at octupole order
and for the 1PN dipole
(Sec.~\ref{sec:res-72}).
These differences reflect the definition of the
mean elements, not the underlying orbital dynamics.

\section{Cross Terms at Orders $\delta\varepsilon^{3/2}$
and $\delta\varepsilon^{7/2}$}
\label{sec:res}

\subsection{Order $\grade{1}{3/2}$}
\label{sec:res-32}

At this order, the only possible contribution is
the second-stage coupling between the two-body
1PN perturbation and the Newtonian quadrupole;
the direct term is absent
(Table~\ref{tab:census}).

After inner-orbit averaging, the two-body 1PN
perturbation reduces to apsidal precession,
\[
  \dot\omega_{\rm 1PN}
  =\frac{3(Gm)^{3/2}}
  {c^2a_1^{5/2}\ell_1^2},
\]
which depends only on $p_1$ and $e_1$ and is
independent of $\M_2$.
Thus, $U_{\rm PN}=0$, and Eq.~\eqref{eq:S2}
gives
\begin{equation}
  \mathsf S_2^{\grade{1}{3/2}}
  =D_X\bar{\bm F}_{\rm PN}
  \cdot\avtw{U_{\rm N}}=0
  \label{eq:32}
\end{equation}
by the zero-time-average condition
\eqref{eq:timegauge}.
Hence, no secular cross term remains at
$\grade{1}{3/2}$ in the time-normalized
mean elements, although periodic corrections
are still present.

Other gauges may introduce removable terms.
The gauge shift $\gamma^\varphi$ contributes
$D_X\bar{\bm F}_{\rm PN}\cdot\gamma^\varphi$,
while the associated clock correction
$\dot\omega_{\rm 1PN}\Delta t_\varphi$
contributes
$-D_X\gamma^\varphi\cdot\bar{\bm F}_{\rm PN}$.
Together, these terms form the commutator in
Eq.~\eqref{eq:gauge}, corresponding to the
removable Hamiltonian term
${\cal K}_F=\{{\cal K}^{\grade{1}{0}},\chi\}$.
For $\varphi=E_2$, both contributions vanish
at quadrupole order
(Sec.~\ref{sec:ms-gauge}).

\subsection{Order $\grade{1}{5/2}$}

Only the direct term appears at this order,
yielding the de Sitter precession of the inner
orbit about the outer orbital angular momentum:
\begin{equation}
  \begin{split}
    \dot{\bm h}
    &=\Omega_{\rm dS}\,
      \hat{\bm z}_2\times\bm h,\\
    \dot{\bm e}_1
    &=\Omega_{\rm dS}\,
      \hat{\bm z}_2\times\bm e_1,
  \end{split}
  \qquad
  \Omega_{\rm dS}
  =\frac{G^{3/2}m_3(4m+3m_3)}
  {2M^{1/2}c^2a_2^{5/2}\ell_2^2},
  \label{eq:dS}
\end{equation}
where $\hat{\bm z}_2$ is the unit normal to
the outer orbital plane.

Using this plane as the reference plane
(Sec.~\ref{sec:frame}), the secular rates are
$\dot\Omega_1=\Omega_{\rm dS}$ and
$\dot p_1=\dot e_1=\dot\iota=\dot\omega_1=0$.
This reproduces the known de Sitter result and
provides a consistency check on the direct term.

\subsection{Order $\grade{1}{7/2}$}
\label{sec:res-72}

Three contributions appear at this order:
the direct term and the first- and second-stage
couplings shown in Fig.~\ref{fig:grade}.

\paragraph*{Direct term $\mathsf D$.}
The direct contribution is obtained by doubly
averaging the $\grade{1}{7/2}$ component of
the EIH relative acceleration.
In the limit $m\ll m_3$, the rates of all five
inner orbital elements
$(p_1,e_1,\iota,\Omega_1,\omega_1)$
agree with those of LR
[their Eqs.~(A7)--(A11)].

The rates of $e_1$ and $\omega_1$ contain
harmonics $(\pm1,\pm1)$ and $(\pm3,\pm1)$
in $(\omega_1,\omega_2)$, whereas the other
rates contain only $(\pm1,\pm1)$.
Taken alone, the direct term produces a nonzero
secular drift of the inner semimajor axis,
\[
  \dot a_1
  =\frac{\dot p_1+2a_1e_1\dot e_1}{\ell_1^2}
  \neq0
\]
(Table~\ref{tab:channels}).

\paragraph*{First stage: Center-of-mass motion $\times$ tide.}
The Newtonian center of mass of the inner binary
moves with velocity
$\bm V_{\rm cm}=-(m_3/M)\dot{\bm R}$.
The individual velocities entering the two-body
EIH terms are
$\bm v_{1,2}=\bm V_{\rm cm}
\pm(m_{2,1}/m)\dot{\bm r}$.
Terms linear in $\bm V_{\rm cm}$ generate a
$\grade{1}{\frac12}$ perturbation proportional
to $m_1-m_2$.

Although its inner average vanishes, its
short-period corrections couple to the
Newtonian quadrupole through Eq.~\eqref{eq:S1},
giving
$\grade{1}{\frac12}+\grade{0}{3}
=\grade{1}{\frac72}$.
Both the orbital-element corrections
($D_X\bm F\cdot u$) and the mean-anomaly
corrections
($\partial_{\M_1}\bm F\,w$)
contribute.

\paragraph*{Second stage: 1PN dipole $\times$ quadrupole.}
After inner-orbit averaging, the
$\grade{1}{2}$ perturbation contains a
1PN dipole term that is odd under
$\bm R\to-\bm R$
\cite{Arakida2026PaperII}:
\begin{equation}
  \bar {\cal H}_{\rm dip}
  =\mathcal C_{\rm dip}
  \frac{\bm e_1\!\cdot\!\bm R}{R^3},
  \qquad
  \mathcal C_{\rm dip}
  =\frac{G^2m_1m_2(m_1-m_2)m_3}
  {4mc^2}.
  \label{eq:dip}
\end{equation}
Its outer-orbit average vanishes.
Using
$\ddot{\bm R}=-GM\bm R/R^3$, the corresponding
periodic generating function is
\begin{equation}
  W_{\rm dip}
  =\int\bar {\cal H}_{\rm dip}\,dt
  =-\frac{\mathcal C_{\rm dip}}{GM}
  \bm e_1\!\cdot\!\dot{\bm R},
  \qquad
  \avt{W_{\rm dip}}=0.
  \label{eq:Wdip}
\end{equation}

This generating function shifts $\bm R$ by
$-Gm_1m_2(m_1-m_2)\bm e_1/(4m^2c^2)$,
corresponding to the orbit-averaged 1PN
displacement of the inner binary's center
of mass.

Because this shift depends on $\bm e_1$,
it also induces the periodic corrections
\[
  \delta\bm j_1
  =\frac{\mathcal C_{\rm dip}}{GML_1}
  \bm e_1\times\dot{\bm R},
  \qquad
  \delta\bm e_1
  =\frac{\mathcal C_{\rm dip}}{GML_1}
  \bm j_1\times\dot{\bm R}.
\]
Their coupling to the Newtonian quadrupole
produces the second-stage contribution
$\mathsf S_2$ at $\grade{1}{\frac72}$.

The 1PN binding-energy contribution coupled
to the Newtonian octupole has the same formal
order but vanishes because
$\bar{\bm F}^{\grade{1}{1}}=0$
(Sec.~\ref{sec:ms-order}).

\paragraph*{Result.}
Combining the three contributions gives
\begin{equation}
  \dot{\bar X}^{\grade{1}{7/2}}
  =\mathsf D+\avtw{\mathsf S_1}+\mathsf S_2
  =\mathsf V_{\Kmq}(\bar X),
  \label{eq:result}
\end{equation}
where $\mathsf V_{\Kmq}$ is the Hamiltonian
vector field generated by the magnetic
1PN--quadrupole Hamiltonian
\begin{equation}
  \begin{split}
    \Kmq={}&
    -\frac{
      3G^2m_1m_2(m_1-m_2)m_3M^{1/2}a_1^{3/2}
    }{
      c^2m^{3/2}a_2^{7/2}\ell_2^5
    }\\
    &\times\big[
      (\bm j_1\times\bm e_1)\!\cdot\!
      (\bm j_2\times\bm e_2)
      -2(\bm j_1\!\cdot\!\bm j_2)
      (\bm e_1\!\cdot\!\bm e_2)
    \big].
  \end{split}
  \label{eq:Kmq}
\end{equation}
Here, $\bm j_i=\ell_i\hat{\bm n}_i$,
and $\bm e_i$ are the eccentricity vectors.
The expression in square brackets is
$e_1e_2\ell_1\ell_2\mathcal B$, where
\begin{equation}
  \mathcal B
  =\cos\iota\cos\omega_1\cos\omega_2
  +(2\cos^2\iota-1)
  \sin\omega_1\sin\omega_2.
  \label{eq:B}
\end{equation}

Eq.~\eqref{eq:Kmq} agrees with the
second term of Eq.~(35) in KST
\cite{KuntzSerraTrincherini2023}
and with the independent canonical result
\cite{Arakida2026PaperII}.
The agreement retains the full mass dependence
and has been verified numerically over
$10^{-4}\le m_3/m\le4\times10^5$
using independent Python and Mathematica
implementations
\cite{ArakidaCodePaperI}.
Each contribution vanishes for
$m_1=m_2$ or $e_2=0$.
Their sum also reproduces $\Kmq$ in the
light-tertiary limit $m_3\ll m$.

Table~\ref{tab:channels} compares the three
contributions for a representative configuration.
The individual terms are comparable to the total
but undergo substantial cancellation:
the sum of their absolute component values may
exceed the corresponding total by a factor of
$4.6$.
The direct term alone violates the Hamiltonian
integrability conditions
(Sec.~\ref{sec:canon-test}), whereas the complete
sum satisfies them.
The semimajor-axis drift produced by the direct
term is canceled by the two coupling terms to
within $10^{-6}$, consistent with a secular
Hamiltonian independent of the mean anomaly.

The decomposition into individual contributions
depends on the variables used.
In canonical variables, the doubly averaged
first-stage contribution vanishes
(Sec.~\ref{sec:canon}).

\begin{table}[tb]
  \caption{
    Contributions to the secular rates of
    the osculating mean elements at
    $\grade{1}{7/2}$, normalized by the
    corresponding components of
    $\mathsf V_{\Kmq}$.
    The last column gives the implied
    semimajor-axis rate $\dot a_1$,
    normalized by the largest component
    of $\mathsf V_{\Kmq}$;
    the complete result has $\dot a_1=0$.
    Parameters:
    $(m_1,m_2,m_3)=(30,20,5000)$,
    $a_1=1$, $e_1=0.4$, $\iota=1$,
    $\omega_1=0.7$, $\Omega_1=0.3$,
    $e_2=0.5$, and $\omega_2=0.9$,
    with $G=c=1$ and the outer orbital
    plane as reference
    (Sec.~\ref{sec:frame}).
    Rates correspond to the
    $\delta\varepsilon^{7/2}$ coefficient.
    All angles are in radians.
  }
  \label{tab:channels}
  \centering\small
  \setlength{\tabcolsep}{4pt}
  \begin{tabular}{@{}lrrrrrr@{}}
    \toprule
    & $\dot p_1$ & $\dot e_1$ & $\dot\iota$
    & $\dot\Omega_1$ & $\dot\omega_1$
    & $\dot a_1$\\
    \midrule
    Direct $\mathsf D$
    & $0.1886$ & $0.2884$ & $0.5869$
    & $0.6392$ & $0.4691$ & $0.0593$\\
    First stage $\avtw{\mathsf S_1}$
    & $-1.8101$ & $-0.6933$ & $-1.0258$
    & $-0.9346$ & $-1.0999$ & $0.6641$\\
    Second stage $\mathsf S_2$
    & $2.6214$ & $1.4050$ & $1.4389$
    & $1.2954$ & $1.6308$ & $-0.7235$\\
    \quad Dip.\ $\to$ quad.
    & $2.6214$ & $1.4193$ & $1.4389$
    & $1.2954$ & $0.4906$ & $-0.7149$\\
    \quad Quad.\ $\to$ dip.
    & $0$ & $-0.0143$ & $0$
    & $0$ & $1.1403$ & $-0.0085$\\
    \midrule
    Total
    & $1.0000$ & $1.0000$ & $1.0000$
    & $1.0000$ & $1.0000$ & $<10^{-6}$\\
    \bottomrule
  \end{tabular}
\end{table}

\paragraph*{Gauge dependence.}
Normalizing the dipole periodic correction
using its eccentric- or true-anomaly average,
rather than its time average, changes the
definition of the mean elements.
Unlike the time average, the corresponding
outer-velocity averages are nonzero:
\begin{equation}
  \begin{split}
    \langle\dot{\bm R}\rangle_E
    &=\frac{n_2a_2e_2}{1+\ell_2}
      \hat{\bm v}_2,\\
    \langle\dot{\bm R}\rangle_F
    &=\frac{n_2a_2e_2}{\ell_2}
      \hat{\bm v}_2.
  \end{split}
\end{equation}
Resubstitution then generates the
gauge-dependent term
$D_X\avtw{\bar{\bm F}_{\rm N}}
\cdot\gamma^\varphi$
at $\grade{1}{\frac72}$.

For the configuration in
Table~\ref{tab:channels}, its magnitude relative
to the corresponding components of
$\mathsf V_{\Kmq}$ ranges from
$0.30$ to $0.92$ for eccentric-anomaly averaging
and from $0.65$ to $1.99$ for true-anomaly
averaging.
Thus, unlike the quadrupole correction, the
dipole correction is sensitive to the averaging
measure.
Secular equations obtained using different
normalizations must therefore be transformed
to a common gauge before comparison.

\subsection{Conditions of Validity}
\label{sec:valid}

The derived coefficients retain the full mass
dependence, without expansion in mass ratios.
This algebraic result should be distinguished
from the validity of the truncated secular
equations, which requires a hierarchical,
weak-field, nonresonant system with
well-separated timescales.

Representative necessary conditions are
\begin{equation}
  \begin{gathered}
    \frac{a_1(1+e_1)}{a_2(1-e_2)}\ll1,
    \qquad
    \kappa\varepsilon^{3/2}\ll1,
    \qquad
    \frac{3\delta}
    {\kappa\varepsilon^{3/2}\ell_1^2}\ll1,\\
    \frac{m_3/m}{\kappa}
    \frac{\varepsilon^{3/2}}{\ell_2^3}\ll1,
    \qquad
    \frac{\delta}{1-e_1}\ll1,
    \qquad
    \frac{\delta(M/m)\varepsilon}
    {1-e_2}\ll1.
  \end{gathered}
  \label{eq:valid}
\end{equation}
These conditions require, respectively,
a convergent multipole expansion,
well-separated orbital periods,
slow 1PN precession and tidal evolution
relative to the outer orbital motion,
and weak gravitational fields at both
pericenters.
For $m_3\gg m$, the fourth condition becomes
particularly restrictive because
$(m_3/m)/\kappa\simeq\kappa$.

The approximation excludes close encounters,
mean-motion resonances, rapid periapsis
precession, and the quasi-secular regime.
High eccentricities may further enhance
individual orbital-element rates.

The angle-based equations assume nonsingular
elements, with
$e_1,e_2\ne0$ and $\sin\iota\ne0$.
The vector expressions in
Eqs.~\eqref{eq:dS} and \eqref{eq:Kmq}
remain regular in these limits.
In particular, $\Kmq=0$ at $e_1=0$, but its
gradient with respect to $\bm e_1$ need not
vanish, so the corresponding vector rates
may remain finite and nonzero.

The extreme mass ratios used in the numerical
tests verify the algebraic coefficients, not
the physical accuracy of the secular
approximation.
Physical applications must also satisfy
Eq.~\eqref{eq:valid}.
We have not directly compared the secular
equations with long-term integrations of the
full EIH three-body system.
Such a comparison would test the accuracy of
the truncated expansion rather than the
algebraic agreement in Eq.~\eqref{eq:result}.
Radiation reaction at 2.5PN order and 2PN
corrections are beyond the scope of this work.

\section{Consistency with Canonical Perturbation Theory}
\label{sec:canon}

\subsection{The Canonical Result}
\label{sec:canon-K}

The secular Hamiltonian is obtained from the
1PN ADM Hamiltonian using Lie--Hori--Deprit
transformations
\cite{Hori1966,Deprit1969,Arakida2026PaperII}.
We use the canonical Jacobi variables of
Eq.~\eqref{eq:legendre} and the Delaunay pairs
$(\ell_1,L_1)$, $(\omega_1,G_1)$, and
$(\Omega_1,H_1)$, where
$L_1=\mu_1(Gma_1)^{1/2}$,
$G_1=L_1\ell_1$, and
$H_1=G_1\cos\iota_1$.
The outer variables are defined analogously
using $\mu_2$ and $M$.

The second-order secular Hamiltonian is
\[
  {\cal K}_2=\left\langle
    {\cal H}_2+\frac12\{{\cal H}_1+{\cal K}_1,W_1\}
  \right\rangle,
\]
where the generating functions are normalized
to have zero time average.

At $\grade{1}{3/2}$, the only possible
Poisson-bracket contribution vanishes:
\begin{equation}
  \avt{\{{\cal K}^{\grade{1}{0}},S\}}
  =-\frac{\partial {\cal K}^{\grade{1}{0}}}{\partial G_1}
  \frac{\partial\avt{S}}{\partial\omega_1}
  =0,
\end{equation}
where ${\cal K}^{\grade{1}{0}}$ is the averaged
two-body 1PN Hamiltonian and $S$ is the
outer-orbit quadrupole generating function.

At $\grade{1}{7/2}$, the three contributions
combine to give
\begin{equation}
  \begin{split}
    {\cal K}^{\grade{1}{7/2}}
    ={}&
    \underbrace{\frac{4m+3m_3}{4M}\Kmq}
    _{\rm direct}
    +\underbrace{0}
    _{\substack{\grade{1}{\frac12}
    \otimes\grade{0}{3}\\
    \text{double average}}}
    +\underbrace{\frac{m_3}{4M}\Kmq}
    _{\rm dipole\otimes quadrupole}
    =\Kmq .
  \end{split}
  \label{eq:Kcan}
\end{equation}

The first-stage contribution is the double average of
\[
  \frac12\{{\cal H}_A+{\cal K}_A,W_B\}
  +\frac12\{{\cal H}_B+{\cal K}_B,W_A\},
\]
where
$A=\grade{1}{\frac12}$,
$B=\grade{0}{3}$,
${\cal K}=\avo{{\cal H}}$, and $W$ denotes the
zero-mean generating function.
The three canonical channels,
$(\ell_1,L_1)$, $(\omega_1,G_1)$,
and $(\Omega_1,H_1)$,
contribute individually, but their combined
double average vanishes.

The dipole--quadrupole contribution follows
from the generating function in
Eq.~\eqref{eq:Wdip}.
The inner-averaged quadrupole Hamiltonian is
\begin{equation*}
  \bar {\cal H}_{\rm Q}
  =CR^{-3}\left[
    \frac{15}{2}
    (\bm e_1\!\cdot\!\hat{\bm N})^2
    -\frac32
    (\bm j_1\!\cdot\!\hat{\bm N})^2
    +\frac12-3e_1^2
  \right],
\end{equation*}
where $C=-G\mu_1m_3a_1^2/2$.

Since $W_{\rm dip}$ depends on the inner orbital
variables only through $\bm e_1$, the contribution
is obtained by averaging the Poisson bracket
of $\bar {\cal H}_{\rm Q}$ and $W_{\rm dip}$ over the
outer orbit, using the vector Poisson brackets
of $(\bm j_1,\bm e_1)$
\cite{TremaineToumaNamouni2009}.

With
$\bm R\times\dot{\bm R}
=(GMp_2)^{1/2}\hat{\bm z}_2$,
the calculation requires two orbital averages:
\begin{equation}
  \begin{split}
    \Big\langle\frac{\hat{\bm N}}{R^4}
    \Big\rangle_t
    (GMp_2)^{1/2}
    &=\frac{n_2e_2}{p_2^2}\hat{\bm u}_2,\\
    \Big\langle\frac{\dot{\bm R}}{R^3}
    \Big\rangle_t
    &=\frac{3n_2e_2}{2p_2^2}\hat{\bm v}_2,
  \end{split}
  \label{eq:twoavg}
\end{equation}
where $\hat{\bm u}_2$ points toward the outer
pericenter and
$\hat{\bm v}_2=\hat{\bm z}_2\times\hat{\bm u}_2$.

The resulting contribution is
\begin{equation}
  \begin{split}
    {\cal K}^{\rm dip\otimes Q}
    ={}&
    -\frac{6C\mathcal C_{\rm dip}n_2e_2}
    {GML_1p_2^2}
    \big[
      (\bm j_1\!\cdot\!\hat{\bm z}_2)
      (\bm e_1\!\cdot\!\hat{\bm u}_2)
    +
      (\bm j_1\!\cdot\!\hat{\bm u}_2)
      (\bm e_1\!\cdot\!\hat{\bm z}_2)
    \big]
    =\frac{m_3}{4M}\Kmq ,
  \end{split}
  \label{eq:Delta2}
\end{equation}
where $\mathcal C_{\rm dip}$ is defined in
Eq.~\eqref{eq:dip}.
The final equality follows from
\begin{equation*}
  \begin{split}
    &(\bm j_1\times\bm e_1)\!\cdot\!
    (\bm j_2\times\bm e_2)
    -2(\bm j_1\!\cdot\!\bm j_2)
    (\bm e_1\!\cdot\!\bm e_2)
    =-\ell_2e_2\big[
      (\bm j_1\!\cdot\!\hat{\bm z}_2)
      (\bm e_1\!\cdot\!\hat{\bm u}_2)
      +(\bm j_1\!\cdot\!\hat{\bm u}_2)
      (\bm e_1\!\cdot\!\hat{\bm z}_2)
    \big].
  \end{split}
\end{equation*}
Both averages in Eq.~\eqref{eq:twoavg}
are proportional to $e_2$.
Thus, the dipole--quadrupole contribution
vanishes for a circular outer orbit.

\subsection{Multiple Scales and Lie Series}
\label{sec:canon-lie}

The agreement between
Eqs.~\eqref{eq:result} and \eqref{eq:Kcan}
can be understood through Lie brackets of
vector fields.

Consider the extended phase space
$z=(X,\phi)$, with $\phi=\phi_1$ in the
first averaging stage and $\phi=\M_2$ in
the second.
The unperturbed flow is
$V_0=(0,n(X))$, where
$n=n_1(X)$ in the first stage and
$n=n_2$ (constant) in the second.

Let $V_A=(\bm F_A,g_A)$ and
$V_B=(\bm F_B,g_B)$ be two perturbing
vector fields, with periodic corrections
$Y_A=(u_A,w_A)$ and $Y_B=(u_B,w_B)$.
With the Lie bracket
\[
  [P,Q]=DQ\cdot P-DP\cdot Q,
\]
the homological equation is
\[
  [V_0,Y_A]
  =V_A-\langle V_A\rangle .
\]
Its phase component contains the clock
correction $D_Xn\cdot u_A$.

The orbital-element component of
$DV_A\cdot Y_B$ is
\[
  (DV_A\cdot Y_B)_X
  =D_X\bm F_A\cdot u_B
  +\partial_\phi\bm F_A\,w_B,
\]
which is precisely the form appearing in the
resubstitution term of Eq.~\eqref{eq:st1sec}.
The homological equations then give
\begin{equation}
  \begin{split}
    \big\langle
      DV_A\cdot Y_B+DV_B\cdot Y_A
    \big\rangle_X
    ={}&
    \big\langle[Y_B,V_A]\big\rangle_X\\
    &+
    \big(
      D\langle Y_B\rangle\cdot\langle V_A\rangle
      +D\langle V_B\rangle\cdot\langle Y_A\rangle
    \big)_X,
  \end{split}
  \label{eq:lie}
\end{equation}
where the subscript $X$ denotes the
orbital-element component.

The clock corrections are already included
in this identity.
Their additional terms lie entirely in the
phase component and therefore do not modify
Eq.~\eqref{eq:lie}.
For a fixed outer orbit, $n_2$ is constant,
so no analogous second-stage clock correction
appears.
Allowing the outer elements to evolve introduces
additional phase corrections, but these do not
affect the inner secular rates at the orders
considered here
(Table~\ref{tab:census}).

In the zero-time-average gauge
\eqref{eq:timegauge}, the terms involving the
mean periodic corrections vanish.
The two resubstitution directions then combine as
\[
  \frac12\big\langle
    [Y_B,V_A]+[Y_A,V_B]
  \big\rangle_X.
\]

In canonical variables, $V_A$ and $Y_A$ are
Hamiltonian vector fields generated by ${\cal H}_A$
and $W_A$, respectively.
The expression above therefore corresponds to
the Poisson-bracket contribution
\[
  \frac12\left\langle
    \{{\cal H}_A,W_B\}+\{{\cal H}_B,W_A\}
  \right\rangle
\]
in the second-order Lie--Hori--Deprit Hamiltonian.

Thus, resubstitution in both directions recovers
the canonical Poisson-bracket contribution.
Each direction separately contains additional
symmetric derivative terms, which cancel in
the sum.
Retaining only one direction may therefore
destroy the Hamiltonian structure.

For osculating elements, the perturbing vector
fields are generally noncanonical because the
1PN acceleration is velocity dependent.
Eq.~\eqref{eq:lie} nevertheless remains a
vector-field identity.
The Hamiltonian consistency of the complete
$\grade{1}{7/2}$ result follows from
Eq.~\eqref{eq:result} together with the
mean-element transformation discussed below.

\subsection{Osculating and Canonical Mean Elements}
\label{sec:canon-mean}

The multiple-scale method uses osculating
elements defined by $(\bm r,\dot{\bm r})$,
whereas canonical perturbation theory uses
contact elements defined by
$(\bm r,\bm p/\mu_1)$
\cite{EfroimskyGoldreich2003,Efroimsky2005}.
At 1PN order,
\[
  \dot{\bm r}-\frac{\bm p}{\mu_1}
  =\frac{1}{c^2}
  \frac{\partial {\cal H}_{\rm 1PN}}{\partial\bm p}.
\]

Three transformations relate the two sets
of mean elements:
the multiple-scale map $T_{\rm osc}$ from
osculating mean elements to osculating elements
[Eq.~\eqref{eq:ansatz}],
the osculating-to-contact map
$C:X\mapsto X+\Gamma$,
and the Lie--Hori--Deprit map $T_{\rm can}$
from canonical mean elements to contact elements.
Here,
\[
  \Gamma
  =-\frac{\partial X}{\partial\dot{\bm r}}
  \cdot\frac{1}{c^2}
  \frac{\partial {\cal H}_{\rm 1PN}}{\partial\bm p}.
\]

Their composition maps osculating mean elements
to canonical mean elements:
\[
  T_{\rm can}^{-1}\circ C\circ T_{\rm osc}
  =\mathrm{id}+\gamma.
\]
Here, $\circ$ denotes composition,
$(A\circ B)(X)=A(B(X))$, so the rightmost
map acts first.
Thus, $T_{\rm osc}$ maps the osculating mean
elements to osculating elements,
$C$ converts them to contact elements,
and $T_{\rm can}^{-1}$ maps the contact elements
to canonical mean elements.

The identity map is denoted by $\mathrm{id}$,
and $\gamma$ is the residual near-identity shift:
\[
  X_{\rm can}
  =X_{\rm osc}+\gamma(X_{\rm osc}).
\]
Each transformation reduces to the identity
when the perturbations vanish, so $\gamma$
measures the difference between the two
definitions of the mean elements.

The corresponding secular vector fields satisfy
\begin{equation}
  \bar{\bm V}_{\rm can}
  =\bar{\bm V}_{\rm osc}
  +[\gamma,\bar{\bm V}]
  +{\cal O}(\gamma^2),
  \label{eq:meanmap}
\end{equation}
where $[\cdot,\cdot]$ is the Lie bracket of
vector fields
(Sec.~\ref{sec:canon-lie}),
generalizing Eq.~\eqref{eq:gauge}.

With the same Newtonian generators and
zero-time-average normalization, the leading
mean-element shift is
$\gamma=\langle\langle\Gamma\rangle\rangle$.
Mean-anomaly shifts do not affect the secular
orbital-element rates because those rates are
independent of the mean anomalies.

The osculating-to-contact shift $\Gamma$ contains
a two-body contribution at $\grade{1}{0}$,
a center-of-mass term at
$\grade{1}{\frac12}$ proportional to the outer
momentum $\bm P$, and three-body terms at
$\grade{1}{1}$ and $\grade{1}{\frac32}$.
Since $\avtw{\bm P}=0$, the
$\grade{1}{\frac12}$ component of $\gamma$
vanishes.
The potentially relevant product of
$\Gamma^{\grade{1}{0}}$ and
$U^{\grade{0}{3}}$ also has zero time average.

A shift $\gamma^{\grade{1}{s}}$ modifies the
1PN secular rates through
$[\gamma^{\grade{1}{s}},
\bar{\bm V}^{\grade{0}{b}}]$
at order $\grade{1}{s+b}$, with $b\ge3$.
A correction at $\grade{1}{\frac72}$ would
therefore require
$\gamma^{\grade{1}{\frac12}}$, which vanishes.

Consequently, the osculating and canonical
mean-element rates agree at
$\grade{1}{\frac72}$.
At the integer orders $\grade{1}{3}$ and
$\grade{1}{4}$, however,
$\gamma^{\grade{1}{0}}$ modifies the rates
through its coupling to the Newtonian quadrupole
and octupole
(Sec.~\ref{sec:canon-test}).
These conclusions apply to the inner orbit
with the outer orbit held fixed.

Although the complete secular rates agree,
their decomposition depends on the choice of
variables.
The first-stage contribution is substantial
in osculating elements
(Table~\ref{tab:channels}), whereas its double
average vanishes in canonical variables
[Eq.~\eqref{eq:Kcan}].

\subsection{Integrability Conditions as a Consistency Test}
\label{sec:canon-test}

With the outer orbit fixed and its plane chosen
as the reference plane
(Sec.~\ref{sec:frame}), a secular Hamiltonian
${\cal K}(L_1,G_1,H_1,\omega_1,\omega_2)$
that is independent of the mean anomaly gives
\[
  \dot L_1=0,\qquad
  \dot G_1=-\frac{\partial {\cal K}}{\partial\omega_1},
  \qquad
  \dot H_1=\frac{\partial {\cal K}}{\partial\omega_2},
\]
together with
\[
  \dot\omega_1=\frac{\partial {\cal K}}{\partial G_1},
  \qquad
  \dot\Omega_1=\frac{\partial {\cal K}}{\partial H_1}.
\]
After elimination of the nodes,
$\partial/\partial\Omega_1
=-\partial/\partial\omega_2$.

Thus, $\dot a_1=0$, and the 1-form
\[
  \dot\omega_1\,dG_1
  +\dot\Omega_1\,dH_1
  -\dot G_1\,d\omega_1
  +\dot H_1\,d\omega_2
\]
must be closed at fixed $L_1$.
This condition yields six integrability
relations, three of which are
\begin{equation}
  \begin{split}
    \frac{\partial\dot G_1}{\partial H_1}
    +\frac{\partial\dot\Omega_1}{\partial\omega_1}
    &=0,\\
    \frac{\partial\dot G_1}{\partial\omega_2}
    +\frac{\partial\dot H_1}{\partial\omega_1}
    &=0,\\
    \frac{\partial\dot\Omega_1}{\partial\omega_2}
    -\frac{\partial\dot H_1}{\partial H_1}
    &=0.
  \end{split}
  \label{eq:tests}
\end{equation}
The remaining three are
\[
  \frac{\partial\dot\omega_1}{\partial H_1}
  =\frac{\partial\dot\Omega_1}{\partial G_1},
  \qquad
  \frac{\partial\dot\omega_1}{\partial\omega_2}
  =\frac{\partial\dot H_1}{\partial G_1},
\]
and the Liouville condition
\[
  \frac{\partial\dot\omega_1}{\partial\omega_1}
  +\frac{\partial\dot G_1}{\partial G_1}=0.
\]

Together with $\dot a_1=0$, these give seven
necessary conditions for canonical
mean-element rates.
They are preserved under canonical transformations
but need not hold for noncanonical mean elements.
Their violation alone therefore does not imply
an inconsistency.

At $\grade{1}{\frac72}$, however, the
transformation between osculating and canonical
mean elements leaves the secular rates unchanged
(Sec.~\ref{sec:canon-mean}), so all seven
conditions apply.
At $\grade{1}{3}$ and $\grade{1}{4}$,
the nonzero $\gamma^{\grade{1}{0}}$ modifies
the osculating mean-element rates, which therefore
need not satisfy all seven conditions.

We checked all seven conditions symbolically
using SymPy and Mathematica.
The Hamiltonian~\eqref{eq:Kmq}, and hence the
complete multiple-scale result
\eqref{eq:result}, satisfies all of them.

The $\grade{1}{3}$ and $\grade{1}{4}$ rates
of LR satisfy Eqs.~\eqref{eq:tests} but violate
some of the remaining conditions.
Both also give $\dot a_1\ne0$, consistent with
their being noncanonical mean-element rates at
these orders.
Their $\grade{1}{\frac72}$ rates, which agree
with the direct term alone, violate all seven
conditions.
Thus, neither the direct term alone nor the
libration term alone is generated by a secular
Hamiltonian in these variables.

These tests concern only the inner orbit with
the outer orbit held fixed.
Conservation of the total angular momentum
$\bm J_1+\bm J_2$ also requires the outer-sector
rates and follows from the rotational invariance
of Eq.~\eqref{eq:Kmq}; it is not tested here.

\section{Summary}
\label{sec:sum}

We have developed a two-stage multiple-scale treatment of the EIH equations
for hierarchical triple systems using osculating elements and time averaging,
with particular emphasis on the 1PN cross terms at
$\delta\varepsilon^{3/2}$ and $\delta\varepsilon^{7/2}$.

With zero-time-average periodic corrections and all relevant couplings included,
no secular cross term remains at $\delta\varepsilon^{3/2}$.
At $\delta\varepsilon^{7/2}$, the complete result reproduces the
inner-orbit secular dynamics generated by the magnetic
1PN--quadrupole Hamiltonian of KST, including its full mass dependence.
This agreement holds within the hierarchical, weak-field, nonresonant regime
specified in Sec.~\ref{sec:valid}.

The $\delta\varepsilon^{7/2}$ contribution consists of three parts:
the direct term, a first-stage coupling between the inner binary's
center-of-mass motion and the Newtonian tidal field, and a second-stage
coupling between the 1PN dipole and the Newtonian quadrupole.
The dipole periodic correction represents the orbit-averaged 1PN displacement
of the inner binary's center of mass.
Although the decomposition into individual contributions depends on the
choice of variables, their sum agrees with the result of
Lie--Hori--Deprit canonical perturbation theory.
Thus, the osculating and canonical formulations yield the same
inner-orbit secular dynamics at this order.

These results highlight four requirements for a consistent
multiple-scale calculation:
\begin{enumerate}[leftmargin=18pt]
\item Use a consistent averaging measure,
      preferably time (or equivalently mean anomaly),
      for all periodic corrections, including clock terms.
\item Retain all contributions at the required perturbative order,
      including resubstitution in both directions and
      mean-anomaly corrections.
\item Compare secular equations using the same definition of the
      mean elements, retain the full mass dependence, and impose
      geometric constraints only after differentiation.
\item Check Hamiltonian integrability conditions where applicable,
      together with known limiting cases.
\end{enumerate}

The discrepancies with LR at
$\delta\varepsilon^{3/2}$ and
$\delta\varepsilon^{7/2}$ arise from different averaging conventions
and omitted coupling terms, respectively.
In particular, all five
$\delta\varepsilon^{7/2}$ rates reported by LR correspond to the
direct contribution alone.
Table~\ref{tab:compare} summarizes the main differences among
previous treatments and the present work.

\begin{table*}[tb]
  \caption{
    Comparison of treatments of 1PN cross terms
    in hierarchical triples.
    ``Mean elements'' denotes the variables
    used in the secular equations.
  }
  \label{tab:compare}
  \centering\small
  \setlength{\tabcolsep}{4pt}
  \begin{tabular}{@{}
    >{\raggedright\arraybackslash}p{0.10\textwidth}
    >{\raggedright\arraybackslash}p{0.18\textwidth}
    >{\raggedright\arraybackslash}p{0.17\textwidth}
    >{\raggedright\arraybackslash}p{0.12\textwidth}
    >{\raggedright\arraybackslash}p{0.15\textwidth}
    >{\raggedright\arraybackslash}p{0.18\textwidth}@{}}
    \toprule
    & Method & Mean elements & Outer orbit
    & $\delta\varepsilon^{3/2}$
    & $\delta\varepsilon^{7/2}$\\
    \midrule
    Will \cite{Will2014,Will2014b}
    & EIH; osculating elements
    & Osculating; time averaging
    & Circular
    & None
    & Vanishes for $e_2=0$\\

    LR \cite{LimRodriguez2020}
    & EIH; two-stage multiple scales
    & Osculating; true-anomaly averaging
      in stage 2
    & Eccentric
    & Libration term
    & Direct term only;
      leading order in $m/M$\\

    KST
    \cite{KuntzSerraTrincherini2021,
    KuntzSerraTrincherini2023}
    & Nonrelativistic EFT
    & Canonical (contact)
    & Eccentric
    & None
    & $\Kmq$\\

    LWYL \cite{LiWuYounsiLi2025}
    & ADM Hamiltonian; von Zeipel
    & Canonical
    & Eccentric
    & No secular term
    & Agreement with KST reported\\

    This work
    & EIH; two-stage multiple scales
      and Lie--Hori--Deprit
    & Osculating and canonical;
      time averaging
    & Eccentric; full mass dependence
    & None
    & $\Kmq$ from three contributions\\
    \bottomrule
  \end{tabular}
\end{table*}

The implications for orbital evolution are examined in the companion paper
\cite{Arakida2026PaperII}, including the disappearance of the
accelerated merger found in LR's octupole example when the libration term
is removed.
\begin{acknowledgments}
The author thanks Halston Lim and Carl L. Rodriguez
for sharing their Mathematica notebooks and C++ code
\cite{LimRodriguezPC}.
This work was supported by a research grant from the
College of Engineering, Nihon University.
Symbolic computations were performed with
Wolfram Mathematica and SymPy, and numerical
computations with NumPy and SciPy \cite{SciPy2020}.
ChatGPT (OpenAI) assisted with English-language
editing, and Claude (Anthropic) with code
development and debugging.
The author independently verified all code
and results and assumes full responsibility
for the article's content.
\end{acknowledgments}
\section*{Author Declarations}

\paragraph*{Conflict of Interest.}
The author declares no conflicts of interest.

\paragraph*{Author Contributions.}
H.A.\ performed all theoretical derivations,
numerical and symbolic computations, and
manuscript preparation.
\section*{Data Availability}

The Python and Mathematica code supporting
this work is publicly available in the repository
cited in Ref.~\cite{ArakidaCodePaperI}
(version 2.2).
The SHA-256 checksum of
\texttt{release\_PaperI\_v2.2.zip} is
\texttt{638e7761731db7edceac2ab4a13e1d28}%
\allowbreak
\texttt{3cdc2495916a0ebb8a7ec83b5a211093}.
The release includes the scripts used for the
comparison with all five $\grade{1}{7/2}$ rates of LR, the symbolic
derivations in Sec.~\ref{sec:ms-order},
the integrability tests in
Sec.~\ref{sec:canon-test}, and the
numerical verification of Eq.~\eqref{eq:result},
including convergence and limiting-case tests.
The \texttt{README} identifies the scripts
used to generate each computed equation, table,
and figure.

\appendix

\section{Time Averages over a Keplerian Orbit}
\label{app:kepler}

Let $e\equiv e_2$, $\ell\equiv\ell_2$, and
$x\equiv(1-\ell)/e=e/(1+\ell)$.
The Fourier expansion
\begin{equation*}
  \frac{1-x^2}{1+x^2+2x\cos F}
  =1+2\sum_{k\ge1}(-x)^k\cos(kF)
\end{equation*}
gives
$\avF{\cos(kF)/(1+e\cos F)}=(-x)^k/\ell$.
Using
\begin{equation*}
  (1+e\cos F)^{-2}
  =(1+e\partial_e)(1+e\cos F)^{-1},
\end{equation*}
we obtain
\begin{equation}
  C_k\equiv\avt{\cos(kF)}
  =(-1)^k(1+k\ell)x^k,
  \label{eq:Ck}
\end{equation}
while $\avt{\sin(kF)}=0$.
The uniform eccentric-anomaly average,
using $d\M_2=(1-e\cos E_2)dE_2$, is
$\langle\cos(kF)\rangle_E=(-x)^k$.

The orbital averages in Eq.~\eqref{eq:twoavg}
follow from
$dt=R^2dF/(GMp_2)^{1/2}$ and
\begin{equation*}
  \dot{\bm R}
  =\left(\frac{GM}{p_2}\right)^{1/2}
  \left[
    -\sin F\,\hat{\bm u}_2
    +(e+\cos F)\hat{\bm v}_2
  \right].
\end{equation*}
These relations also give
$\langle\dot{\bm R}\rangle_E$ and
$\langle\dot{\bm R}\rangle_F$
used in Sec.~\ref{sec:res-72}.

\section{The Libration Term of LR}
\label{app:lib}

\subsection{The Term and Its Origin}

We distinguish the equations of LR
from the Mathematica and C++ implementations
provided by the authors \cite{LimRodriguezPC}.
Our main conclusions follow from the published
equations; implementation-specific observations
are identified separately.

Eq.~(4.5) of LR introduces a secular
contribution to the inner argument of
pericenter at $\grade{1}{3/2}$.
In their implementation, this term is
\begin{equation}
  \Big(\frac{d\omega_1}{dt}\Big)_{\rm LR}
  =-\frac{15\,G^{3/2}m\,m_3\,e_1^2}
  {4c^2\sqrt M\,a_1a_2^{3/2}
  \,\ell_1^3\ell_2^3}
  \,C_2\,\Theta,
  \label{eq:LR45}
\end{equation}
where
$C_2=(1-\ell_2)(1+2\ell_2)/(1+\ell_2)$
and
\begin{equation*}
  \begin{split}
    \Theta={}
    \cos\iota\cos2\omega_1\cos2\omega_2
    +\frac12(1+\cos^2\iota)
    \sin2\omega_1\sin2\omega_2.
  \end{split}
\end{equation*}
The published expression differs from
Eq.~\eqref{eq:LR45} by a factor of
$-\ell_1\ell_2^3$.

After inner-orbit averaging, LR use the
outer true anomaly $F$ as the independent
variable and impose zero uniform
$F$ average on the periodic corrections
[their Eq.~(3.38)].

The libration term arises from substituting
the quadrupole periodic correction into
the two-body 1PN apsidal precession rate,
as in Eq.~\eqref{eq:S2}.
Since $\dot\omega_{\rm 1PN}$ depends only
on $p_1$ and $e_1$, and the quadrupole
conserves $a_1$, this gives
\begin{equation}
  \Big(\frac{d\omega_1}{dt}\Big)^{\rm resub}
  =\frac{\dot\omega_{\rm 1PN}}{\ell_1^2}
  \,\avt{W_{e_1^2}},
  \label{eq:idnq}
\end{equation}
where $W_{e_1^2}$ is the periodic
correction to $e_1^2$.
The time average differs from the
uniform $F$ average because the
Keplerian factor $dt/dF$ is nonconstant.

\subsection{The Averaging-Measure Mismatch in Closed Form}
\label{app:lib-closed}

The inner-averaged quadrupole gives
$d(e_1^2)/dt$ proportional to
$R^{-3}(\bm e_1\!\cdot\!\hat{\bm N})
\hat{\bm N}\!\cdot\!(\bm e_1\times\bm j_1)$.
The corresponding periodic correction is
\begin{equation}
  W_{e_1^2}
  =\mathcal C\left[
    \frac{\sin^2\iota}{4}\sin2\omega_1
    (F+e_2\sin F-\M_2)+W_T
  \right]+\gamma,
  \label{eq:Wsol}
\end{equation}
where
$\mathcal C
=15m_3a_1^{3/2}e_1^2\ell_1/
[(mM)^{1/2}p_2^{3/2}]$,
and $W_T$ contains harmonics
$2\omega_2+kF$ with $k=1,2,3$.

LR choose $\gamma$ so that
$\avF{W_{e_1^2}}=0$, whereas
Eq.~\eqref{eq:idnq} requires its
time average.
The nontrigonometric term in
Eq.~\eqref{eq:Wsol} has the same
average under both measures.
The remaining difference is
\begin{equation*}
  \avt{W_T}-\avF{W_T}
  =\Lambda(\alpha\sin2\omega_2
  -\beta\cos2\omega_2),
\end{equation*}
where
$\Lambda=\tfrac12C_2+\tfrac{e_2}{2}C_1
+\tfrac{e_2}{6}C_3=-C_2/6$
and
$\alpha\sin2\omega_2-\beta\cos2\omega_2
=\Theta/2$.
Consequently,
\begin{equation}
  \avt{W^{\rm LR}_{e_1^2}}
  =\avt{W_{e_1^2}}-\avF{W_{e_1^2}}
  =-\frac{C_2}{12}\mathcal C\Theta.
  \label{eq:ce2}
\end{equation}
Substitution into Eq.~\eqref{eq:idnq}
reproduces Eq.~\eqref{eq:LR45},
including its sign and all factors
of $\ell_1$ and $\ell_2$.

The eccentricity dependence is governed by
$C_2=\avt{\cos2F}$, which vanishes
as $e_2\to0$.
Time and true-anomaly averages coincide
in this limit, explaining the absence
of the term for circular outer orbits
\cite{Will2014,Will2014b}.
Under zero-time-average normalization,
the libration contribution vanishes
[Eq.~\eqref{eq:32}].

\subsection{The Missing Companion Terms}
\label{app:lib-KF}

The change in averaging normalization
can be represented by a canonical
transformation generated by
$\chi=\avt{S}-\avF{S}$, where
\begin{equation}
  \begin{split}
    \chi={}&
    \frac{\sqrt G\,\mu_1m_3a_1^2}
    {24\sqrt M\,p_2^{3/2}}\,C_2
    \Big\{
      \Big[
        \frac{15}{2}e_1^2
        (\cos^2\omega_1
        -\sin^2\omega_1\cos^2\iota)\\
        &+\frac32\ell_1^2\sin^2\iota
      \Big]\sin2\omega_2
      -15e_1^2\cos\omega_1\sin\omega_1
      \cos\iota\cos2\omega_2
    \Big\}.
  \end{split}
  \label{eq:chi}
\end{equation}
In the $F$ gauge, the corresponding
Hamiltonian term is
\begin{equation}
  {\cal K}_F^{\grade{1}{3/2}}
  =\{{\cal K}^{\grade{1}{0}},\chi\}
  =\frac{15\,G^2\mu_1m^{3/2}m_3}
  {8c^2\sqrt M\,a_1^{1/2}p_2^{3/2}}
  \,C_2\,\frac{e_1^2}{\ell_1^2}\,\Theta .
  \label{eq:KF}
\end{equation}

Libration term of LR corresponds to
$-2{\cal K}_F/G_1$, one contribution to
$\dot\omega_1=\partial {\cal K}_F/\partial G_1$.
The remaining components of the
Hamiltonian flow are absent from reported rates of LR.
For example, ${\cal K}_F$ also generates
\begin{equation}
  \dot e_1
  =\frac{15\,G^{3/2}m\,m_3\,e_1}
  {8c^2\sqrt M\,a_1\ell_1\,p_2^{3/2}}
  \,C_2\,
  \frac{\partial\Theta}{\partial\omega_1}.
  \label{eq:e1comp}
\end{equation}

In a consistent $F$ gauge, these
companion terms arise when the 1PN
clock correction
$\dot\omega_{\rm 1PN}\Delta t_F$
is resubstituted into the quadrupole
rates.
Together with the original
resubstitution, they form the commutator
in Eq.~\eqref{eq:gauge}.

In the implementation provided by LR,
however, the clock correction uses
the outer eccentric anomaly,
$\dot\omega_{\rm 1PN}\Delta t_E$,
for which the corresponding
resubstitution vanishes
(Sec.~\ref{sec:ms-gauge}).
Thus, the two coupling channels use
different averaging conventions.
The resulting rates do not correspond
to a single definition of the
mean elements.

\subsection{Consistency Tests}

Because ${\cal K}$ is rational in the
Delaunay momenta, the Laurent expansion
of $\partial {\cal K}/\partial G_1$ cannot
contain a $G_1^{-1}$ term.
Both the published and implemented forms of Eq.~(4.5) of LR contain such
a term in one angular component
of $\Theta$.
Neither form, taken alone, can therefore
arise from a Hamiltonian rational
in the Delaunay momenta.

The Liouville residual is
\begin{equation}
  \begin{split}
    \frac{\partial\dot\omega_1}{\partial\omega_1}
    +\frac{\partial\dot G_1}{\partial G_1}
    &=-\frac{15\,G^{3/2}m\,m_3\,e_1^2e_2^2
    (1+2\ell_2)}
    {4c^2\sqrt M\,p_1p_2^{3/2}
    \,\ell_1(1+\ell_2)^2}\\
    &\quad\times\big[
      (1+\cos^2\iota)
      \cos2\omega_1\sin2\omega_2
      -2\cos\iota
      \sin2\omega_1\cos2\omega_2
    \big],
  \end{split}
  \label{eq:resid}
\end{equation}
which is canceled by the companion
contribution to $\dot G_1$.

\begin{figure*}[tb]
  \centering
  \includegraphics[width=0.9\linewidth]{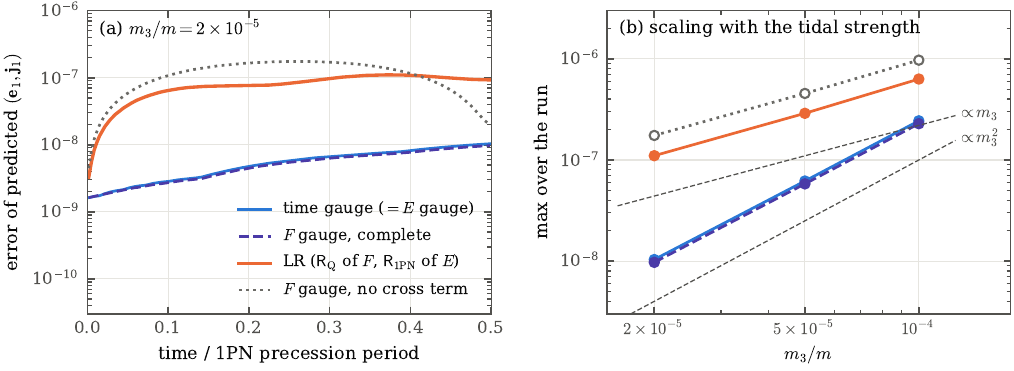}
  \caption{
    Comparison of averaging gauges using
    inner-averaged dynamics with 1PN
    apsidal precession and the Newtonian
    quadrupole perturbation.
    The tertiary follows a fixed outer orbit
    with $e_2=0.9$ and
    $\dot\omega_{\rm 1PN}P_{\rm out}
    =2\times10^{-3}$.
    (a) Errors in the reconstructed
    osculating orbit for
    $m_3/m=2\times10^{-5}$.
    (b) Maximum error versus $m_3/m$.
    Errors scale as $m_3^2$ for
    consistent gauges and as $m_3$
    for combination and the $F$ gauge without companion terms of LR.
  }
  \label{fig:phase}
\end{figure*}

Figure~\ref{fig:phase} compares
averaging gauges using the inner-averaged
equations with a fixed outer orbit.
This test concerns only the
$\grade{1}{\frac32}$ coupling between
1PN apsidal precession and the
Newtonian quadrupole, not the
$\grade{1}{\frac72}$ terms or
the full EIH dynamics.

The time, $E_2$, and complete $F$
gauges reproduce the reference orbit
with errors of second order in the
tidal perturbation, whereas 
combination produces first-order errors of LR.

Two additional implementation issues
are the imposition of geometric
constraints before differentiation
and the omission of the
$\Delta\Omega$ contribution.
Neither affects the 1PN--quadrupole
cross terms, since the two-body
1PN acceleration has no out-of-plane
component and $\dot\omega_{\rm 1PN}$
is independent of $\iota$.
Their effects arise at
quadrupole-squared order or higher.

\section{The $\delta\varepsilon^{7/2}$ Terms of LR}
\label{app:72}

In  implementation by LR, the
$\grade{1}{7/2}$ rates contain only
the direct contribution obtained by doubly
averaging the three-body EIH acceleration.
The first- and second-stage resubstitution
terms are absent.

These rates correspond to the terms
proportional to $e_2p_2^{-7/2}$ in
Eqs.~(A7)--(A11) of LR.
The rates of $p_1$, $\iota_1$, and
$\Omega_1$ are
\begin{equation}
  \begin{split}
    \dot p_1
    &=\frac14\mathcal S p_1^2
    \big[
      \cos\iota(5\cos^2\iota-1)
      \sin\omega_1\cos\omega_2
      +(3-7\cos^2\iota)
      \cos\omega_1\sin\omega_2
    \big],\\
    \dot\iota_1
    &=\frac{1}{16}\mathcal S p_1\sin\iota
    \big[
      (16-10\cos^2\iota)
      \sin\omega_1\cos\omega_2
      +14\cos\iota
      \cos\omega_1\sin\omega_2
    \big],\\
    \dot\Omega_1
    &=\frac18\mathcal S p_1
    \frac{\sin\iota}{\sin\iota_1}
    \big[
      7\cos\omega_1\cos\omega_2
      +17\cos\iota
      \sin\omega_1\sin\omega_2
    \big],
  \end{split}
  \label{eq:LR72}
\end{equation}
where
$\mathcal S
=3e_1e_2G^{3/2}M^{3/2}\ell_2^3
(1-4\nu)^{1/2}/
(c^2p_2^{7/2}\ell_1^2)$.

LR also provide the rates of $e_1$
and $\omega_1$ [their Eqs.~(A8) and (A10)].
These contain harmonics
$(\pm1,\pm1)$ and $(\pm3,\pm1)$
in $(\omega_1,\omega_2)$ and can be
written as
\[
  \dot e_1
  =-\frac{\mathcal S\ell_1^2p_1}
  {8e_1(1+\ell_1)^2}\mathcal E,
  \qquad
  \dot\omega_1
  =-\frac{\mathcal S p_1}
  {8e_1^2(1+\ell_1)^2}\mathcal W,
\]
where $\mathcal E$ and $\mathcal W$
are the angular polynomials given by LR.

We evaluated the direct contribution
$\mathsf D$ in Sec.~\ref{sec:res-72}
without expanding in mass ratios.
In the limit $m\ll m_3$, it reproduces
all five rates of LR. For ten random configurations at
$m_3/m=10^7$, the discrepancy is below
$3\times10^{-6}$, consistent with
numerical accuracy.
The comparison accounts for 
orientation convention of LR for $\bm e_1$,
which is opposite to that adopted here.

Thus, direct average of LR is consistent
with the leading-order large-tertiary-mass
limit of our result.
The factor $M^{3/2}$ represents this
leading mass dependence, while finite
mass-ratio corrections remain in
$\mathsf D$.
The osculating direct term also differs
from the canonical direct contribution
$(4m+3m_3)\Kmq/(4M)$
[Eq.~\eqref{eq:Kcan}], because the
decomposition depends on the variables
(Sec.~\ref{sec:canon-mean}).

Since rates of LR coincide with
$\mathsf D$, they also produce a
nonzero secular drift of the
osculating semimajor axis,
\[
  \dot a_1
  =\frac{\dot p_1+2a_1e_1\dot e_1}
  {\ell_1^2}\ne0.
\]
For ten random configurations,
$|\dot a_1|$ ranges from $0.7\%$
to $32\%$ of the largest
orbital-element rate.

The discrepancy with KST arises
because LR omit the first- and
second-stage coupling terms identified
in Sec.~\ref{sec:res-72}.
These contributions can exceed the
direct term in magnitude.
Their inclusion restores agreement
with the magnetic 1PN--quadrupole
Hamiltonian, as shown in
Eq.~\eqref{eq:result}.

Consistent with this incomplete
Hamiltonian structure,  $\grade{1}{7/2}$ rates of LR
violate all seven integrability conditions
(Sec.~\ref{sec:canon-test}).
The difference in angular dependence
from KST, including terms proportional
to $\sin^2\iota$, is likewise resolved
by the omitted couplings.
\section{Osculating, Contact, and Mean Elements}
\label{app:elements}

Osculating elements are defined by the
Keplerian orbit determined from
$(\bm r,\dot{\bm r})$.
They obey the Gauss planetary equations
with the perturbing acceleration
in Eq.~\eqref{eq:relEOM}
and are used by LR and in
Secs.~\ref{sec:ms} and \ref{sec:res}.

For velocity-dependent perturbations,
the canonical momentum differs from
$\mu_1\dot{\bm r}$.
The Keplerian elements defined by
$(\bm r,\bm p/\mu_1)$ are called
contact elements
\cite{EfroimskyGoldreich2003,Efroimsky2005}.
Their Delaunay variables are canonical
and are used by KST, LWYL,
and our canonical formulation.

At 1PN order, the two sets of elements
are related by
\begin{equation}
  X^{\rm con}
  =X^{\rm osc}
  -\frac{\partial X}{\partial\dot{\bm r}}
  \cdot\frac{1}{c^2}
  \frac{\partial {\cal H}_{\rm 1PN}}{\partial\bm p}
  +{\cal O}(c^{-4}).
\end{equation}

Mean elements are obtained by removing
short-period variations.
Their definition depends on the averaging
normalization
(Sec.~\ref{sec:ms-gauge}).
The time-normalized osculating mean
elements and canonical mean elements
are related by the averaged
osculating-to-contact transformation
[Eq.~\eqref{eq:meanmap}].
This transformation does not modify the
secular rates at $\grade{1}{7/2}$
(Sec.~\ref{sec:canon-mean}).

Table~\ref{tab:conv} summarizes the
conventions used by LR, KST, and this work.
KST define contact elements in the
inner binary's rest frame, whereas our
canonical elements use the Jacobi momentum
$\bm p$ in the zero-total-momentum frame.
The difference is a boost associated
with the center-of-mass motion, whose
1PN correction is linear in the outer
momentum $\bm P$.
Its time average vanishes, so it does
not affect the $\grade{1}{7/2}$
secular rates
(Sec.~\ref{sec:canon-mean}).

In comparing $\Kmq$ with KST,
we account for the opposite outer-vector
convention and the sign relating their
averaged Lagrangian to our Hamiltonian.
The Mathematica script
\texttt{PaperI\_Cross72}
reproduces the second term of KST's
Eq.~(35) under these conventions.

\begin{table*}[tb]
  \caption{
    Conventions used by LR
    \cite{LimRodriguez2020},
    KST \cite{KuntzSerraTrincherini2023},
    and this work.
  }
  \label{tab:conv}
  \centering\small
  \begin{tabular}{@{}
    >{\raggedright\arraybackslash}p{0.17\textwidth}
    >{\raggedright\arraybackslash}p{0.26\textwidth}
    >{\raggedright\arraybackslash}p{0.26\textwidth}
    >{\raggedright\arraybackslash}p{0.24\textwidth}@{}}
    \toprule
    & This work & LR & KST\\
    \midrule
    Outer vector
    & $\bm R=\bm x_3-\bm X_{\rm N}$
      (binary to tertiary)
    & Same as this work
    & $\bm R=\bm Y_{\rm CM}-\bm y_3$
      (tertiary to binary)\\

    Inner eccentricity vector
    & Toward the pericenter of
      $\bm r=\bm x_1-\bm x_2$
    & Opposite orientation for
      $m_1>m_2$ (heavier to lighter)
    & ---\\

    Orbital elements
    & Osculating ($\dot{\bm r}$);
      canonical in Sec.~\ref{sec:canon}
    & Osculating ($\dot{\bm r}$)
    & Contact ($\bm p$);
      vectors $\bm j_i$, $\bm e_i$\\

    Reference plane
    & Outer orbital plane
      (outer orbit fixed)
    & Invariable plane;
      $\iota=\iota_1+\iota_2$,
      $\Omega_2=\Omega_1+\pi$
    & Frame-independent vectors\\

    Mean elements
    & Zero time average in both stages
    & Zero uniform $F$ average in
      stage 2 [their Eq.~(3.38)]
    & Near-identity canonical
      transformations\\

    $\grade{1}{7/2}$ contribution
    & Hamiltonian $\Kmq$
      [Eq.~\eqref{eq:Kmq}]
    & Rates in Eqs.~(A7)--(A11)
    & Second term of Eq.~(35);
      averaged Lagrangian,
      $\Kmq=-{\cal L}$\\

    Mass dependence
    & Full mass dependence
    & $(1-4\nu)^{1/2}M^{3/2}$;
      leading order in $m/M$
    & Full mass dependence\\
    \bottomrule
  \end{tabular}
\end{table*}
\section{Numerical Order Separation}
\label{app:num}

We fix $a_1$ and set $a_2=a_1\lambda^{-2}$.
The outer position and velocity become
\begin{equation*}
  \bm R=a_1\lambda^{-2}\bm\rho(\M_2),
  \qquad
  \dot{\bm R}
  =\lambda\left(\frac{GM}{a_1}\right)^{1/2}
  \bm\nu(\M_2),
\end{equation*}
where $\bm\rho$ and $\bm\nu$ describe
a Keplerian orbit with unit semimajor axis.
The distances to the tertiary can be written as
\begin{equation*}
  r_{A3}
  =a_1\lambda^{-2}
  \left|
    \bm\rho-\lambda^2\beta_A
    \frac{\bm r}{a_1}
  \right|,
\end{equation*}
with $\beta_1=m_2/m$ and
$\beta_2=-m_1/m$.
This makes the perturbing acceleration
analytic in $\lambda$ near $\lambda=0$.

For complex $\lambda$, the norm denotes
the analytic continuation
$|\bm z|=(\bm z\cdot\bm z)^{1/2}$,
using the nonconjugated scalar product
and the branch positive for real $\lambda$.
The radicand remains away from the branch
cut on the contours used ($|\lambda|\le0.3$).

The coefficient of $\lambda^k$ corresponds
to $\grade{n}{k/2}$ in the $n$PN sector.
Python extracts these coefficients using
Cauchy's integral formula on contours
of radius $0.12$--$0.2$, sampled at
$32$--$48$ points; Mathematica uses
\texttt{Series}.
The $c^{-2}$ contributions come from
the 1PN part of Eq.~\eqref{eq:EIH}
and the bilinear couplings in
Eqs.~\eqref{eq:S1}--\eqref{eq:S2}.

\section{Second-Order Terms in the Extended Phase Space}
\label{app:ext}

We derive Eq.~\eqref{eq:lie}.
Let $z=(X,\phi)$, $V_0=(0,n(X))$, and
$V_A=\langle V_A\rangle+\tilde V_A$,
where $\langle\cdot\rangle$ denotes
averaging over $\phi$ at fixed $X$.
The periodic correction $Y_A$ satisfies
the homological equation
\begin{equation}
  [V_0,Y_A]
  =DY_A\cdot V_0-DV_0\cdot Y_A
  =\tilde V_A,
  \label{eq:homol}
\end{equation}
or, in components,
\[
  n\,\partial_\phi u_A=\tilde{\bm F}_A,
  \qquad
  n\,\partial_\phi w_A
  =\tilde g_A+D_Xn\cdot u_A,
\]
as in Eq.~\eqref{eq:st1per}.

Because $V_0$ has only a phase component,
$DV_0\cdot Z$ and $D^2V_0[Z,Z']$
also have only phase components.
Moreover, for any periodic $Z$,
\[
  \langle DZ\cdot V_0\rangle
  =n\langle\partial_\phi Z\rangle=0.
\]
We use $\simeq$ to denote equality
of the averaged $X$-components.

Substituting
$V_A=\langle V_A\rangle+[V_0,Y_A]$
and the corresponding expression for
$V_B$, we obtain
\begin{equation*}
  \begin{split}
    \langle DY_B\cdot V_A
    +DV_B\cdot Y_A\rangle
    &\simeq
    D\langle Y_B\rangle
    \cdot\langle V_A\rangle
    +D\langle V_B\rangle
    \cdot\langle Y_A\rangle\\
    &\quad+
    \big\langle
      DY_B\cdot[V_0,Y_A]
      +D[V_0,Y_B]\cdot Y_A
    \big\rangle.
  \end{split}
\end{equation*}
The last integrand can be written as
\begin{equation*}
  \begin{split}
    DY_B\cdot[V_0,Y_A]
    +D[V_0,Y_B]\cdot Y_A
    &=D(DY_B\cdot Y_A)\cdot V_0\\
    &\quad-D^2V_0[Y_B,Y_A]
    -DV_0\cdot DY_B\cdot Y_A.
  \end{split}
\end{equation*}
Its average has no $X$-component:
the first term averages to zero,
and the remaining terms affect only
the phase.
This establishes Eq.~\eqref{eq:lie}.
In particular, the clock corrections
involving $D_Xn$ and $D_X^2n$
do not contribute to the averaged
$X$-equations.

For the symmetry used in
Sec.~\ref{sec:canon-lie},
the Jacobi identity and
Eq.~\eqref{eq:homol} give
\begin{equation*}
  \begin{split}
    [Y_B,V_A]-[Y_A,V_B]
    ={}&[Y_B,\langle V_A\rangle]
       -[Y_A,\langle V_B\rangle]
      -[V_0,[Y_A,Y_B]].
  \end{split}
\end{equation*}
The last term has zero averaged
$X$-component.
The remaining terms give
\begin{equation*}
  \begin{split}
    D\langle V_A\rangle
    \cdot\langle Y_B\rangle
    -D\langle Y_B\rangle
    \cdot\langle V_A\rangle
    -(A\leftrightarrow B),
  \end{split}
\end{equation*}
which vanishes in the time gauge.

At the second stage, $n=n_2$
is constant, so $DV_0=0$ and
the same argument applies without
clock corrections.
\section{Intermediate Results of the Canonical Calculation}
\label{app:canint}

We summarize intermediate results of the
canonical calculation in Sec.~\ref{sec:canon-K},
using the notation of this paper with $G=c=1$.
The Mathematica script \texttt{PaperI\_Cross72}
also verifies by exact simplification that
${\cal H}^{\grade{1}{1/2}}$ and
${\cal H}^{\grade{1}{7/2}}$ agree with the
corresponding terms of the ADM Hamiltonian
in \texttt{PaperI\_LieHori}.
Here $\mathcal B$ is defined in Eq.~\eqref{eq:B},
while $F$ and $\omega_2$ denote the outer
true anomaly and argument of pericenter.

\paragraph*{Dipole generating function.}
The zero-time-average antiderivative
of $\bar {\cal H}_{\rm dip}$ with respect to
$\M_2$ [Eq.~\eqref{eq:Wdip}] is
\begin{equation}
  \begin{split}
    \mathcal A_{\rm dip}
    ={}&\frac{e_1m_1m_2(m_1-m_2)m_3}
    {4m\,a_2^2\ell_2}
    \big[
      \cos\iota\,
      (e_2\cos\omega_2+\cos(F+\omega_2))
      \sin\omega_1\\
    &\quad-\cos\omega_1\,
      (e_2\sin\omega_2+\sin(F+\omega_2))
    \big].
  \end{split}
\end{equation}
The script verifies that
$\partial\mathcal A_{\rm dip}/\partial\M_2
=\bar {\cal H}_{\rm dip}$
and $\avt{\mathcal A_{\rm dip}}=0$.

\paragraph*{Canonical contributions.}
Writing Eq.~\eqref{eq:Kmq} as
\begin{equation*}
  \Kmq
  =-\frac{
    3m_1m_2(m_1-m_2)m_3M^{1/2}
    a_1^{3/2}e_1\ell_1e_2
  }{
    m^{3/2}a_2^{7/2}\ell_2^4
  }\,\mathcal B,
\end{equation*}
the direct contribution and the
dipole--quadrupole coupling are
$(4m+3m_3)\Kmq/(4M)$ and
$m_3\Kmq/(4M)$, respectively.
Their sum is $\Kmq$
[Eq.~\eqref{eq:Kcan}].

The direct term is obtained analytically
by reducing ${\cal H}^{\grade{1}{7/2}}$
to inner-orbit moments and direction
cosines. The reduction is also
checked numerically against the
original Hamiltonian.

\paragraph*{First-stage contribution.}
The first-stage term is
\[
  \Delta_1
  =\frac12\left(
    \langle X_1\rangle+\langle X_2\rangle
  \right),
\]
where
\[
  X_1=\{{\cal H}_A+{\cal K}_A,W_B\},
  \qquad
  X_2=\{{\cal H}_B+{\cal K}_B,W_A\}.
\]
The Poisson brackets include the
three conjugate pairs
$(\ell_1,L_1)$, $(\omega_1,G_1)$,
and $(\Omega_1,H_1)$;
$\langle\cdot\rangle$ denotes
the double average.

For
$e_1=3/10$, $e_2=1/2$,
$a_1=1$, $a_2=12$,
$(m_1,m_2,m_3)=(2,1,3/2)$,
$\iota=1$, $\omega_1=2/5$,
and $\omega_2=7/5$,
the three contributions to
$\langle X_1\rangle$ are
$-3.6280\times10^{-6}$,
$2.9723\times10^{-6}$, and
$6.5568\times10^{-7}$.
The corresponding contributions to
$\langle X_2\rangle$ are identical.
They cancel in the sum.

The analytic verification uses
\texttt{TrigReduce} to expand
$\sin\iota\,\langle X_{1,2}\rangle$
in harmonics of $\iota$,
$\omega_1$, and $\omega_2$.
All 33 Fourier coefficients for
each bracket simplify exactly to
zero, establishing
$\Delta_1=0$.
\bibliography{refsI}

\end{document}